\documentclass[a4paper,11pt]{article}
\pdfoutput=1
\usepackage{jcappub}
\usepackage[T1]{fontenc}
\usepackage[utf8]{inputenc}
\usepackage{mathrsfs}
\usepackage{bm}
\usepackage{url}
\usepackage[normalem]{ulem}
\usepackage{mathtools}
\usepackage{array}
\newcolumntype{P}[1]{>{\centering\arraybackslash}p{#1}}
\newcolumntype{M}[1]{>{\centering\arraybackslash}m{#1}}
\usepackage[caption=false]{subfig}
\usepackage{dcolumn}
\usepackage{graphicx, epsfig}
\usepackage[dvipsnames]{xcolor}
\usepackage{mathrsfs}
\usepackage{bm}
\usepackage{yhmath}
\usepackage[caption=false]{subfig}
\usepackage[normalem]{ulem}
\usepackage{mathtools}
\usepackage{bigints}
\usepackage{float}
\usepackage{hyperref}
\def\dms{\hbox{$\Delta m^2$ }}

\def\t13{\mathrel{{\theta_{13}}}}
\def\y12{\mathrel{{\tan^2 \theta_{12}}}}
\def\c2{\mathrel{{\chi^2 }}}

\newcommand{\be}{\begin{equation}}
\newcommand{\ee}{\end{equation}}
\newcommand{\ba}{\begin{eqnarray}}
\newcommand{\ea}{\end{eqnarray}}

\title{
On probing turbulence in core-collapse supernovae in upcoming neutrino detectors
}
\author[a,1]{Mainak Mukhopadhyay,\note{Corresponding author.}}
\author[b]{and Manibrata Sen}
\affiliation[a]{Department of Physics; Department of Astronomy \& Astrophysics; Center for Multimessenger Astrophysics, Institute for Gravitation and the Cosmos, The Pennsylvania State University, University Park, PA 16802, USA.}
\affiliation[b]{Max-Planck-Institut für Kernphysik,
Saupfercheckweg 1, 69117 Heidelberg, Germany.}
\emailAdd{mkm7190@psu.edu}
\emailAdd{manibrata@mpi-hd.mpg.de}
\abstract{
Neutrino propagation through a turbulent medium can be highly non-adiabatic leading to distinct signatures in the survival probabilities. A core-collapse supernova can be host to a number of hydrodynamic instabilities which occur behind the shockfront. Such instabilities between the forward shock and a possible reverse shock can lead to cascades introducing turbulence in the associated matter profile, which can imprint itself in the neutrino signal. In this work, we consider realistic matter profiles and seed in the turbulence using a randomization scheme to study its effects on neutrino propagation in an effective two-flavor framework. We focus on the potential of upcoming neutrino detectors - DUNE and Hyper-Kamiokande to constrain the parameters characterizing turbulence in a supernova. We find that  these experiments can effectively constrain the parameter space for the amplitude of the spectra, they will only have mild sensitivity to the spectral index, and cannot inform on deviations from the usual Kolmogorov $5/3$ inverse power law. Furthermore, we also confirm that the double-dip feature, originally predicted in the neutrino spectra associated with forward and reverse shocks, can be completely washed away in the presence of turbulence, leading to total flavor depolarization. 
}
\begin{document}
\maketitle
\flushbottom
\section{Introduction}
\label{sec:intro}
The detection of neutrinos from SN1987A can be regarded as one of the first instances of multi-messenger astronomy, where neutrino events were observed hours before visible light. Although limited in statistics - a total of $\mathcal{O}(30)$ neutrino events in the Kamiokande detector~\cite{Kamiokande-II:1987idp,Hirata:1988ad}, the IMB detector~\cite{Bionta:1987qt} and the Baksan detector~\cite{Alekseev:1987ej} were observed - this data has been used widely by the particle and astroparticle physics community to probe important questions of fundamental physics.

It is already well-known that neutrinos from a galactic core-collapse supernova (SN) will be our best bet to probe the supernova dynamics. These neutrinos originate from the deepest regions of the SN and propagate through the entire matter envelope before exiting the SN and arriving at the Earth. The flavour composition of the neutrinos is extremely sensitive to the background fermion density it encounters en route. Close to the neutrinosphere, the neutrino density is large enough, so neutrino forward scattering on a neutrino background is the dominant contribution to the propagation Hamiltonian. This can lead to collective flavour conversion - both ``slow''~\cite{Duan:2005cp,Hannestad:2006nj,Duan:2007mv,PhysRevD.83.105022,Fogli_2007,Raffelt:2007yz,Esteban-Pretel:2007jwl,Dasgupta:2007ws,Dasgupta:2008my, Dasgupta:2010cd, Raffelt:2007cb,Sarikas:2011am, Banerjee:2011fj, Raffelt:2013rqa,Chakraborty:2015tfa, Mirizzi:2012wp,Capozzi:2016oyk, Das:2017iuj,Airen:2018nvp,Stapleford:2019yqg} and  '`fast''~\cite{ Sawyer:2015dsa, Chakraborty:2016lct,Dasgupta:2016dbv,Sawyer:2005jk,Sawyer:2008zs,Sawyer:2015dsa,Chakraborty:2016lct,Dasgupta:2016dbv,Izaguirre:2016gsx,Capozzi:2017gqd, Dighe:2017sur, Capozzi:2017gqd, Dasgupta:2017oko,Dasgupta:2018ulw,Abbar:2018shq,Azari:2019jvr,Johns:2019izj,Glas:2019ijo,Shalgar:2019qwg,Abbar:2020fcl,Bhattacharyya:2020dhu,Bhattacharyya:2020jpj,Li:2021vqj,Padilla-Gay:2021haz,Wu:2021uvt,Richers:2021nbx,Richers:2021xtf,Martin:2019gxb,Abbar:2018beu,Johns:2020qsk,Sigl:2021tmj,Xiong:2021dex,Abbar:2021lmm,DelfanAzari:2019tez,Chakraborty:2019wxe,Capozzi:2020kge,Bhattacharyya:2021klg,Nagakura:2021suv,Nagakura:2021hyb,Dasgupta:2021gfs,Shalgar:2021wlj,Dighe:2017sur,Bhattacharyya:2022eed,Wu:2017drk,Zaizen:2019ufj,Nagakura:2019sig}  (see \cite{Duan:2010bg,Mirizzi:2015eza,Tamborra:2020cul,Richers:2022zug} for detailed reviews on the subject)- which is the major mechanism of flavour transformation within radii of $\sim\mathcal{O}(100)\,$km from the neutrinosphere. 

At larger radii, neutrino density falls off and matter effects, due to coherent forward scattering on background electrons, become important. Multiple studies have demonstrated the sensitivity of the neutrino spectra to the underlying SN density profile through the Mikheyev-Smirnov-Wolfenstein (MSW) resonances~\cite{Mikheev:1986gs,PhysRevD.17.2369}, including signatures of the shockwave propagation at late times~\cite{Schirato:2002tg,Takahashi:2002yj,Lunardini:2003eh,Fogli:2003dw,Fogli:2004ff,Tomas:2004gr,Dasgupta:2005wn,Galais:2009wi}. In the absence of the shockwave, neutrino propagation is shown to be adiabatic, governed only by the vacuum mixing angle and the matter density. However, violations of adiabaticity can arise due to an abrupt change in density at the shockfront. This non-adiabaticity leaves its imprint on the flavour transition and can be used to reconstruct a real-time propagation of the shockwave. 

\begin{figure}
\centering
\includegraphics[width=0.75\textwidth]{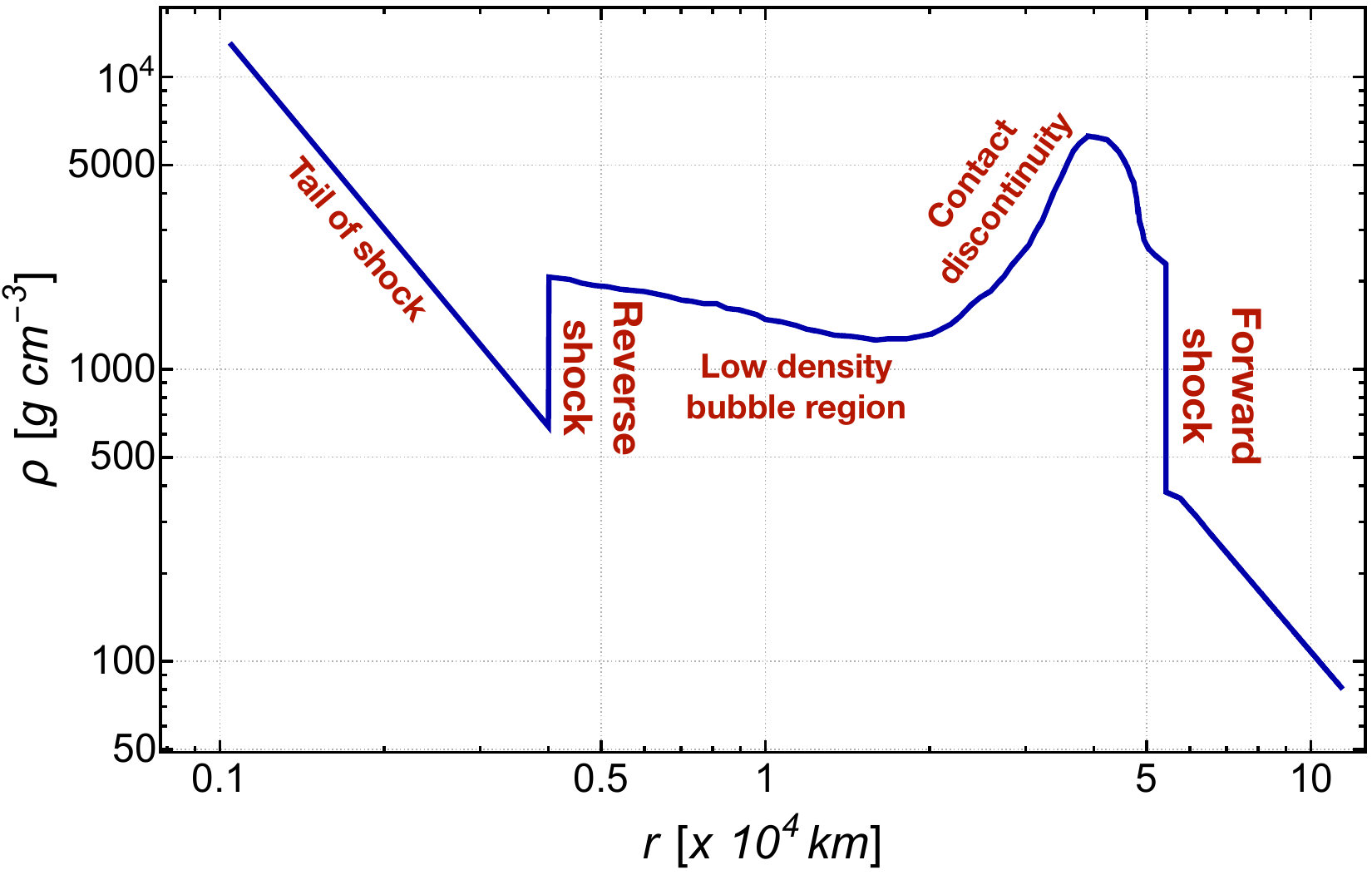}
\caption{\label{fig:density_profile} The density profile snapshot at time $t=5$ s post collapse~\cite{Tomas:2004gr}. The various regions of interest in the shock wave profile are labelled.
}
\end{figure}

A distinctive signature of the non-adiabaticity introduced by the shockwave propagation can manifest in the form of dips in the energy-dependent survival probabilities of neutrinos. A few seconds post the SN core bounce, a possible reverse shockwave can arise if a supersonic neutrino-driven wind collides with the ejecta, as shown in Fig.\,\ref{fig:density_profile}. Note that this is very sensitive to whether the neutrino-driven wind is supersonic or subsonic, for a detailed discussion, see~\cite{Friedland:2020ecy}. 
In \cite{Tomas:2004gr}, it was demonstrated that for neutrinos passing through the two density discontinuities, characterised by the reverse and the forward shocks, a ``double-dip'' feature would be present in the neutrino survival probability as a function of energy. Such a signal can be present in electron neutrinos for normal mass ordering (NMO) and electron antineutrinos for inverted mass ordering (IMO) and can be used to trace the position of the two shockfronts. 

However, this is not the complete story. A SN progenitor is host to a variety of hydrodynamic instabilities, which take place behind the shockfront. This leads to turbulence in the ejecta, causing stochastic fluctuations in the matter density. Turbulence is generally expected to be present in aspherical simulations of SNe, and can play an important role in seeding the explosion~\cite{Radice:2017kmj}. These stochastic fluctuations in the matter density have a non-trivial impact on neutrino flavour evolution.

Neutrino flavour evolution in a turbulent medium has been studied extensively to determine its impact on the neutrino spectra from the next Galactic supernova~\cite{Fogli:2006xy,friedland2006neutrino,Kneller:2010sc,Borriello:2013tha, Lund:2013uta,Kneller:2014oea,Patton:2014lza,Yang:2015oya,Kneller:2017lqg,Abbar:2020ror}. Generically, the effect of turbulence on flavour propagation can be broadly classified into two types: (i) `stimulated' or non-adiabatic transitions between different propagation eigenstates if the stochastic density fluctuations are present near the MSW resonance region~\cite{Kneller:2010sc}, and (ii) phase effects due to the presence of semi-adiabatic resonances~\cite{Dasgupta:2005wn}. The latter effect is more subtle and can be suppressed if the amplitude of fluctuations is large. 

It is numerically challenging to obtain realistic turbulent matter density profiles from aspherical (3D) hydrodynamic simulations. Theoretically, the effect of turbulence has been modelled by adding disturbances to a smooth matter profile obtained from coarse, spherically symmetric simulations. This can be done in a number of different ways - for example, by adding delta correlated noise~\cite{Fogli:2006xy}, or by modelling the turbulent density fluctuations through a Kolmogorov-type power spectrum~\cite{friedland2006neutrino}. In particular, recent 2D hydrodynamic simulations have shown the presence of a power-law dependence of the power spectrum on the wavelength of fluctuations. The long-wavelength fluctuation indeed shows a power-law dependence with the mean spectral index given by $p=1.85^{+0.54}_{-0.77}$, thus indicating that 2D turbulence might be closer to having a Kolmogorov-type power spectrum~\cite{Borriello:2013tha}. However, simulations also indicate that the turbulence might not be fully developed, hence other values of the spectral indices cannot be ruled out. Moreover, it is well known that the cascading of turbulence in 3D is different from 2D~\cite{Dolence:2012kh,Couch:2014kza}. However, the lack of sufficient resolution makes it difficult to recover the Kolmogorov power law behaviour from 3D simulations.

A theoretical study of the impact of turbulence on the neutrino survival probability was performed in~\cite{friedland2006neutrino}, where power-law fluctuations in the density profile were assumed. It was found that the electron neutrino survival probability exhibits a power law when the amplitude of turbulence is increased and a corresponding analytical estimate to explain the power law for the simple turbulence model was derived. This was followed by a series of works~\cite{Patton:2014lza,Yang:2015oya,Kneller:2017lqg}, where the authors used rotating wave approximation to predict the final survival probability amongst other techniques both numerical and analytical. The latter formalism was also used to predict the dependence of the neutrino flavour transition on the amplitude and the spectral index of turbulence.
A theoretical study of the impact of turbulence on the neutrino survival probability was performed in~\cite{friedland2006neutrino}, where power-law fluctuations in the density profile were assumed. It was found that the electron neutrino survival probability exhibits a power law when the amplitude of turbulence is increased and a corresponding analytical estimate to explain the power law for the simple turbulence model was derived. This was followed by a series of works~\cite{Patton:2014lza,Yang:2015oya,Kneller:2017lqg}, where the authors used rotating wave approximation to predict the final survival probability amongst other techniques both numerical and analytical. The latter formalism was also used to predict the dependence of the neutrino flavor transition on the amplitude and the spectral index of turbulence.
However, certain open issues remain. For example, a major unanswered question is whether future-generation neutrino experiments designed to detect thousands of neutrinos from the next galactic supernova actually distinguish between different parameters characterising the turbulence in a supernova. This is the major goal of our work. Furthermore, we aim to shed some light on the presence of the questions ``double-dip'' feature pointed out in~\cite{Tomas:2004gr}.

In this work, we perform a detailed effective two-flavor study of neutrino propagation through a typical SN media profile, characteristic of the cooling phase. We add turbulence to the smooth density profile following the prescription in~\cite{Lund:2013uta}. We find that the addition of turbulence with an amplitude beyond $10\%$ washes out the double-dip feature completely, thereby confirming the results in \cite{Fogli:2003dw,Kneller:2007kg}. We simulate neutrino events in the upcoming DUNE and Hyper-Kamiokande experiment and find that they can distinguish between a turbulent and a non-turbulent SN matter profile with high significance. On the other hand, they do not have enough sensitivity to probe the index of the power spectrum.

Our paper is organized as follows. We discuss turbulence in supernovae and the slab-approximation used as the numerical technique in Sec.~\ref{sec:turb_sn}. The algorithm to seed the turbulence along with its impact on neutrino propagation is discussed in Sec.~\ref{sec:numerics}. The effects of turbulence on the double-dip feature in the neutrino spectra are studied in Sec.~\ref{sec:dbldip}. In Sec.~\ref{sec:evrates}, we discuss the event rates in DUNE and Hyper-Kamiokande along with their potential to constrain the physical parameters associated with turbulence for a galactic scale supernova. We conclude and discuss the implications of our work in Sec.~\ref{sec:concl}. The numerical algorithm developed for this work is summarized in Appendix.~\ref{appsec:numerical_tech}.
\section{Turbulence in supernovae}
\label{sec:turb_sn}
We work in the approximate framework of 2-neutrino oscillations, where we have the electron neutrinos ($\nu_e$) and $\nu_x$ where, $x = \{\mu, \tau\}$, along with each of their antineutrino components. The Equation of Motion (EoM) for a neutrino flavour $|\nu_\alpha\rangle$ can be written as
\begin{equation}
    \mathrm{i}\frac{d}{dt}|\nu_\alpha\rangle= \left[U H_{\rm vac} U^\dagger + V(r) \right] |\nu_\alpha\rangle\,,
    \label{eq:EoM}
\end{equation}
where $H_{\rm vac}={\rm diag}(\Delta m^2/(2E),-\Delta m^2/(2E))$ is the vacuum Hamiltonian and $U$ is the $2\times2$ unitary mixing matrix. The matter potential $V(r)$ has a contribution from the forward scattering of neutrinos off the background fermions and is given by $V_0(r)=\sqrt{2}G_F n_e(r){\rm diag}(1,0)$, where $G_F$ is the Fermi constant and $n_e(r)$ is the electron number density at a given distance $r$. This is extracted from a snapshot of the matter density at a certain time from a simulation. We further neglect the contributions from the non-electron flavour neutrinos, which cancel at the tree-level and arise only at the loop-level~\cite{PhysRevD.35.896, Akhmedov:2002zj,Mirizzi:2009td}. Neutrinos close to the neutrinosphere also experience a forward scattering potential due to their scattering on other background neutrinos~\cite{Pantaleone:1994ns}. This potential is given by $V_{\nu\nu}\simeq\sqrt{2} G_F n_{\nu}$, where $n_{\nu}$ is the total background neutrino density. This neutrino-neutrino potential dominates deep inside the SN, and can lead to fascinating collective effects. However, in the region of our interest, close to the shockfront, this neutrino potential is subdominant, and collective oscillations can be safely ignored.

The turbulence is included in our model by adding a random field $F(r)$ as
\be
\label{eq:effpot}
V(r) = \big(1 + F(r) \big) V_0 (r)
\ee
The random field $F(r)$ is defined in the following way
\be
\label{eq:rand_field1}
F(r) = \begin{cases}
C_* \tanh{\bigg( \frac{r-r_{\rm rs}}{\lambda} \bigg)} \tanh{\bigg(\frac{r_{\rm fs} - r}{\lambda} \bigg)} \times F_{\rm rand} (r),\hspace{0.2cm} r_{\rm rs} \leq r \leq r_{\rm fs}\,,\\
0, \hspace{0.2cm} \text{elsewhere}\,,
\end{cases}
\ee
where $C_*$ is the amplitude of the fluctuations, $F_{\rm rand}(r)$ is a real-valued homogenous Gaussian scalar random field. The reverse shock position is given by $r_{\rm rs}$ while the forward shock position is $r_{\rm fs}$. The $\tanh$ terms suppress the fluctuations close to the shocks and prevent discontinuities at $r_{\rm rs}$ and $r_{\rm fs}$. The parameter $\lambda$ is the length scale over which the fluctuations reach their extent. We choose $\lambda = 100$ km.

The standard way of calculating the evolution of the neutrino wavefunctions involves numerically solving the Schr$\ddot{\rm o}$dinger equation directly. Unfortunately for irregular potentials (matter densities), this method is not efficient. In particular, the current scenario has turbulences making the matter potential highly noisy and the standard scenario for solving such problems is unsuitable. Hence, we use the \emph{slab approximation}~\cite{Giunti:2007ry} to calculate the spatial evolution of the neutrino wavefunctions.

In this approximation, the density profile is approximated by a series of slabs of constant density. We discretize the spatial dimension of physical size $L$ into $N$ slabs, thus the width of each slab is, $\Delta x = L/N$. In such a discretization scheme, the starting point is given by $x_0$ and the boundaries of the slabs are at $x_1,x_2,\dots, x_n$. The mass eigenstates of the neutrinos propagate as plane waves in constant density regions with phases given by ${\rm exp}{(\pm i \Delta m_M^2 \Delta x/4 E)}$, where the effective mass-squared difference in matter is given by
\be
\label{eq:deltamsq}
\Delta m^2_M = \sqrt{\Big( \dms \cos{(2 \theta)} - A_{CC} \Big)^2 + \Big( \dms \sin{(2 \theta)}\Big)^2}\,,
\ee
with $\theta$ being the vacuum mixing angle. 
$A_{CC}$ is the effective charge current potential,
\be
\label{eq:effprof}
A_{CC}(x) = 2 \sqrt{2} E G_F n_e(x)\,.
\ee
All the information about the matter density profile is provided by this term. The wavefunction in the flavour basis is given by,
\be
\bm \Psi_{e} = \langle x|\nu_e\rangle =\begin{pmatrix}
\psi_{ee} \\
\psi_{ex}
\end{pmatrix}\,.
\ee
The wavefunction at the boundaries is joined according to the following scheme

\ba
\label{eq:maineq}
\bm \Psi_e(x_n) &= \left[ U_M\ \mathcal{U}_M(x_n - x_{n-1})\ U_M^\dagger \right]_{(n)} \left[ U_M\ \mathcal{U}_M(x_{n-1} - x_{n-2})\ U_M^\dagger \right]_{(n-1)} \dots \nonumber\\
&\left[ U_M\ \mathcal{U}_M(x_1 - x_0)\ U_M^\dagger \right]_{(1)} \bm \Psi_e(x_0)\,,
\ea
where, 
\begin{itemize}
\item the unitary evolution operator in each slab diagonal basis $\mathcal{U}_M (\Delta x)$ is given by,
\be
\label{eq:unitev}
\mathcal{U}_M (\Delta x) = {\rm diag} \Bigg( {\rm exp}{\big(+ i \Delta m_M^2 \Delta x/4 E\big)},exp{\big(- i \Delta m_M^2 \Delta x/4 E \big)} \Bigg)\,,
\ee
\item $U_M$ is the unitary  mixing matrix in matter given by
\be
\label{eq:unitar}
U_M=\begin{pmatrix}
\cos{ \theta_M} & \sin{ \theta_M} \\
-\sin{ \theta_M} & \cos{ \theta_M}
\end{pmatrix}\,,
\ee
where the effective mixing angle in matter $\theta_M$ is
\be
\label{eq:thetam}
\tan{(2 \theta_M)} = \frac{\tan{(2 \theta)}}{1-\bigg( A_{CC} / \big(\Delta m^2 \cos{(2 \theta)}\big)\bigg)}\,.
\ee
\item The notation $\left[ \dots \right]_{(i)}$ indicates that all the matter-dependent quantities inside the brackets need to be evaluated with the matter density profile of the $i$-th slab, where the slab goes from $x_{i-1}$ to $x_i$.
\end{itemize}
The wavefunction in the matter basis can be easily obtained using~\eqref{eq:maineq} and~\eqref{eq:unitar} as
\be
\label{eq:masswvfn}
\bm \Phi = \begin{pmatrix}
\phi_{1} \\
\phi_{2}
\end{pmatrix}
= U_M^{-1} \bm \Psi_e\,.
\ee
The probability that a neutrino of flavour $\alpha$ is found in some flavour $\beta$ after passing through the matter profile is given by
\be
\label{eq:prob}
P_{\nu_\alpha \rightarrow \nu_\beta} = |\psi_{\alpha \beta}|^2\,.
\ee
Similarly, the corresponding survival probability of the matter eigenstate is given by $P_{\nu_i \rightarrow \nu_j} = |\bm \Phi_i|^2 $. The adiabaticity parameter, which is an important quantity to understand the changes in oscillation and transition probabilities related to neutrino propagation, is defined as~\cite{Giunti:2007ry}
\be
\label{eq:adia}
\gamma = \frac{\Delta m^2_M}{4 E |d \theta_M/dx|} = \frac{(\Delta m_M^2)^2}{2 E\ ({\rm sin}\ 2\theta_M)\ |d A_{{\rm CC}}/dx|}\,.
\ee

Armed with this, we can investigate the propagation of neutrinos in an irregular potential.
\section{Turbulence: Numerical ansatz}
\label{sec:numerics}
In this section, we discuss the Randomization Method used to implement stochastic fluctuations following a power-law on top of a smooth matter profile. 
\subsection{The Randomization Method: Generating the turbulence}
\label{subsec:turb_ansatz}
In previous works, the authors of~\cite{Kneller:2010sc} used the \emph{Randomization Method}~\cite{randomization} to generate and sample the turbulence. In~\cite{Lund:2013uta}, a slightly advanced version of the randomization method - \emph{`Variant C'} was used. For this work, we will use the Variant C of the randomization method to generate the turbulence. It is important to note here that in this method the modes are distributed in a logarithmic scale which makes the distribution scale-invariant by ensuring that the realizations are uniform over the decades of length scales considered in the work. We summarize the method below.

Let the wavenumber space from which we draw the wavenumbers have a size $\Delta$, such that, $\Delta = [0,\infty)$. For numerical implementation, the space $\Delta$ is divided into $n$ non-intersecting intervals $\Delta_j$, such that, $\Delta = \cup_{j=1}^n \Delta_j$. We then select $n_0$ independent random wavenumbers from each of these intervals $\Delta_j$, which can now be indexed as $k_{jl}$, where, $l=1,2,\dots,n_0$. The selection of the $k_{jl}$'s are according to the probability distribution function $p_j (k)$ defined as
\be
\label{eq:prob_k}
p_j(k) = 
\begin{cases}
2 E(k)/\sigma_j^2, \hspace{5mm} \text{for }k \in \Delta_j\,,\\
0, \hspace{5mm} \text{for } k \notin \Delta_j\,,
\end{cases}
\ee
where $E(k)$ is the power spectrum, which we assume to be of the following form,
\be
\label{eq:EK}
E(k) = (\alpha-1)\bigg( \frac{k_*}{k} \bigg)^\alpha\,.
\ee
In the above definition of the power spectrum, $\alpha$ is the spectral index, $k_* = 2 \pi/\lambda_*$ is the wavenumber cutoff, i.e., the cutoff that decides the longest non-zero turbulence wavelength ($\lambda_*$) and hence the smallest wavenumber. Thus, it can be referred to as the driving scale for our turbulence model~\cite{Kneller:2014oea}. For this work, we treat the spectral index $\alpha$ as a free parameter that is varied and the corresponding sensitivity of the next-generation neutrino detectors is tested. We choose, $\lambda_* = 2 (r_{\rm fs} - r_{\rm rs})$, which is twice the distance between the positions of the reverse and forward shocks. In Eq.~\eqref{eq:prob_k}, $\sigma_j^2 = 2 \int_{\Delta_j} E(k) dk$.

\begin{figure}[!t]
\centering
\includegraphics[width=0.75\textwidth]{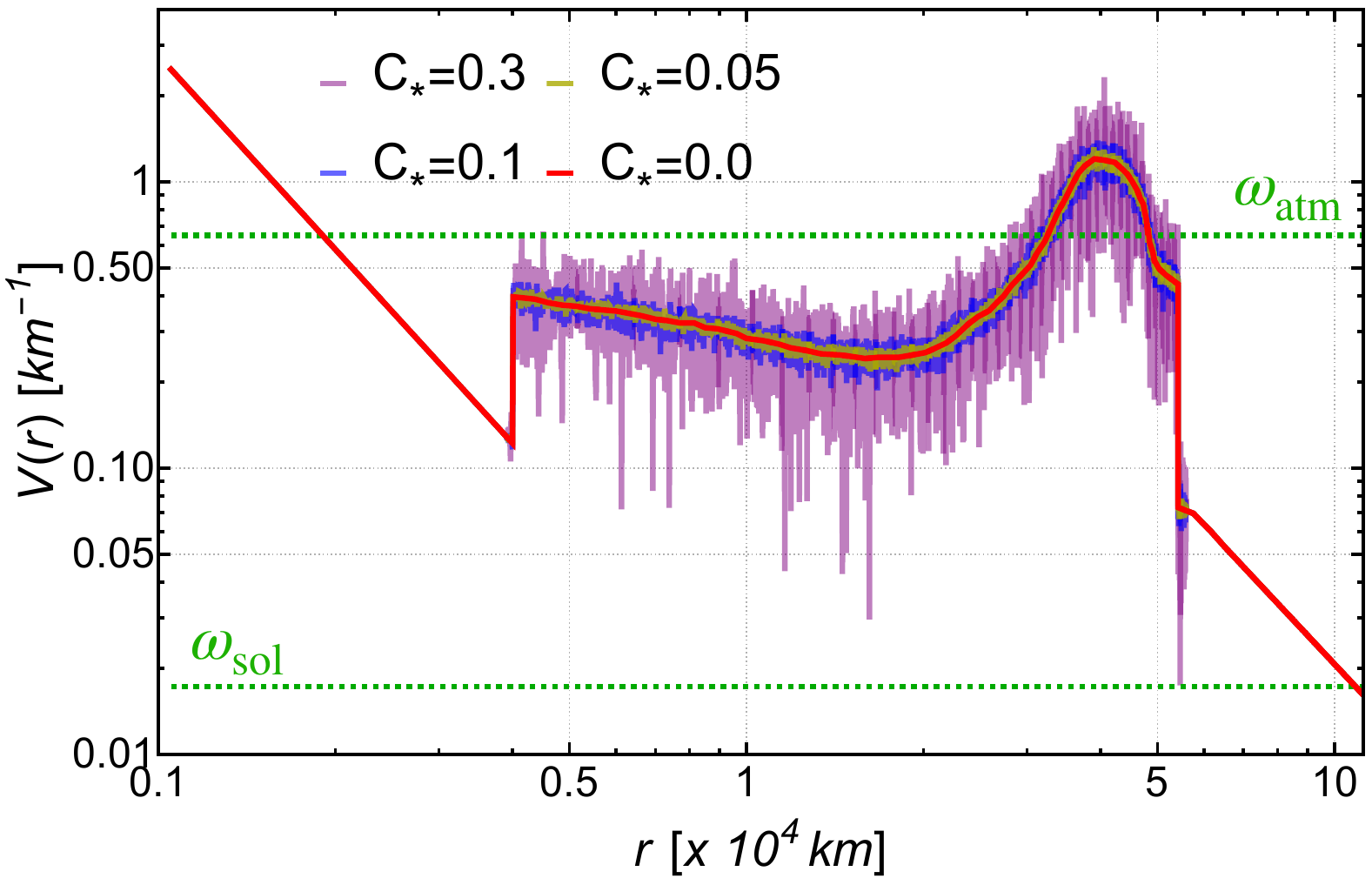}
\caption{\label{fig:turbprofs} The effective matter potential profile for fixed $\alpha = -5/3$ and different values of the normalization $C_*$. The case where turbulence is absent is represented by $C_* = 0$ (solid red curve). The value of $\omega = \Delta m^2/(2 E_\nu)$ is also shown as a dashed green line corresponding $\Delta m^2_{\rm atm} = 2.6 \times 10^{-3}\ \rm eV^2$, $\Delta m^2_{\rm sol} = 6.9 \times 10^{-5}\ \rm eV^2$ for $E_\nu = 10\ \rm MeV$.
}
\end{figure}
We can now define $F_{\rm rand} (r)$ (see Eqs.~\eqref{eq:effpot} and~\eqref{eq:rand_field1}) as
\be
F_{\rm rand} (r) = \sum_{j=1}^n \frac{\sigma_j}{\sqrt{n_0}} \sum_{l=1}^{n_0} \left[ \xi_{jl}\ \text{cos} \big( 2 \pi k_{jl} r \big) + \eta_{jl}\ \text{sin} \big( 2 \pi k_{jl} r \big) \right]\,,
\ee
where, $\xi_{jl}$ and $\eta_{jl}$ are standard Gaussian random variables, which are independent of each other and of $k_{jl}$, have zero mean, and unit variance. Note that the indices $j= 1,2,\dots,n$ and $l=1,2,\dots,n_0$. In this strategy of stratified sampling of the wavenumber bins, we choose the sampling bins such that they are equally spaced in the log-space, that is with respect to $\text{ln } k$. Thus we have, for a given index $j$, $\hat{k}_{j+1} = q \hat{k}_j$, where $q$ is some ratio parameter of the geometric distribution which defines the wavenumber intervals $\Delta_j = (\hat{k}_j,\hat{k}_{j+1}]$. Since we have a minimum wavelength in our model, we have $\hat{k}_1 = k_*$. The bounded domain for $k$ also introduces the constraint, $k\leq k_{\max}$, where we choose $k_{\max} = 100$, such that $\hat{k}_{n+1} = k_{\max}$.
For this work, we chose $n=4$ and $n_0 = 40$. We have also verified that pushing $n$ and $n_0$ to higher values of $5$ and $50$ respectively does not change the results. It is important to note here that this strategy of stratified sampling is efficient in way that allows us to work with a comparatively low value of total number of modes, $\mathcal{O}(100)$, to give us accurate and physical scale independent results, as compared to just sampling the modes trivially, thus, gaining some computational advantage. The density profile extends from $r_{\rm start} = 0.105 \times 10^4$ km to $r_{\rm end} = 11.35 \times 10^4$ km. This region is sampled in spatial steps of $\Delta x = 10$ km. The turbulence is introduced to this density profile, starting at the reverse shock position which is at, $r_{\rm rs} = 0.39 \times 10^4$ km and it is terminated at the position of the forward shock, $r_{\rm fs} = 5.6 \times 10^4$ km. The region of turbulence has the same spatial discretization scale as the original density profile ($= 10$ km). 

As stated earlier, the primary parameters that we vary and examine in this work are the amplitude associated with the turbulence $C_*$ and the power spectral index $\alpha$. In Fig.~\ref{fig:turbprofs}, we show the effective matter potential for a fixed $\alpha = -5/3$ and various values of $C_*$. The red thick line is the case when turbulence is absent, that is, $C_* = 0$. This is essentially the same curve as shown in Fig.~\ref{fig:density_profile} (solid blue curve) rescaled to convert the matter density to effective potential using natural units $\hbar$ and $c$. As the amplitude of turbulence increases, the fluctuations grow as expected. The effects of this variation in the amplitude on the survival probabilities for neutrinos are discussed in the following section. 
It is important to note here that we want to focus on the effects associated solely with the non-adiabaticity at the forward and reverse shocks, hence we artificially smoothed the contact discontinuity so as to avoid any non-adiabaticity effects arising from it.
\subsection{Results}
\label{subsec:turb_results}
\begin{figure}
\subfloat[(a)] {\includegraphics[width=0.49\textwidth]{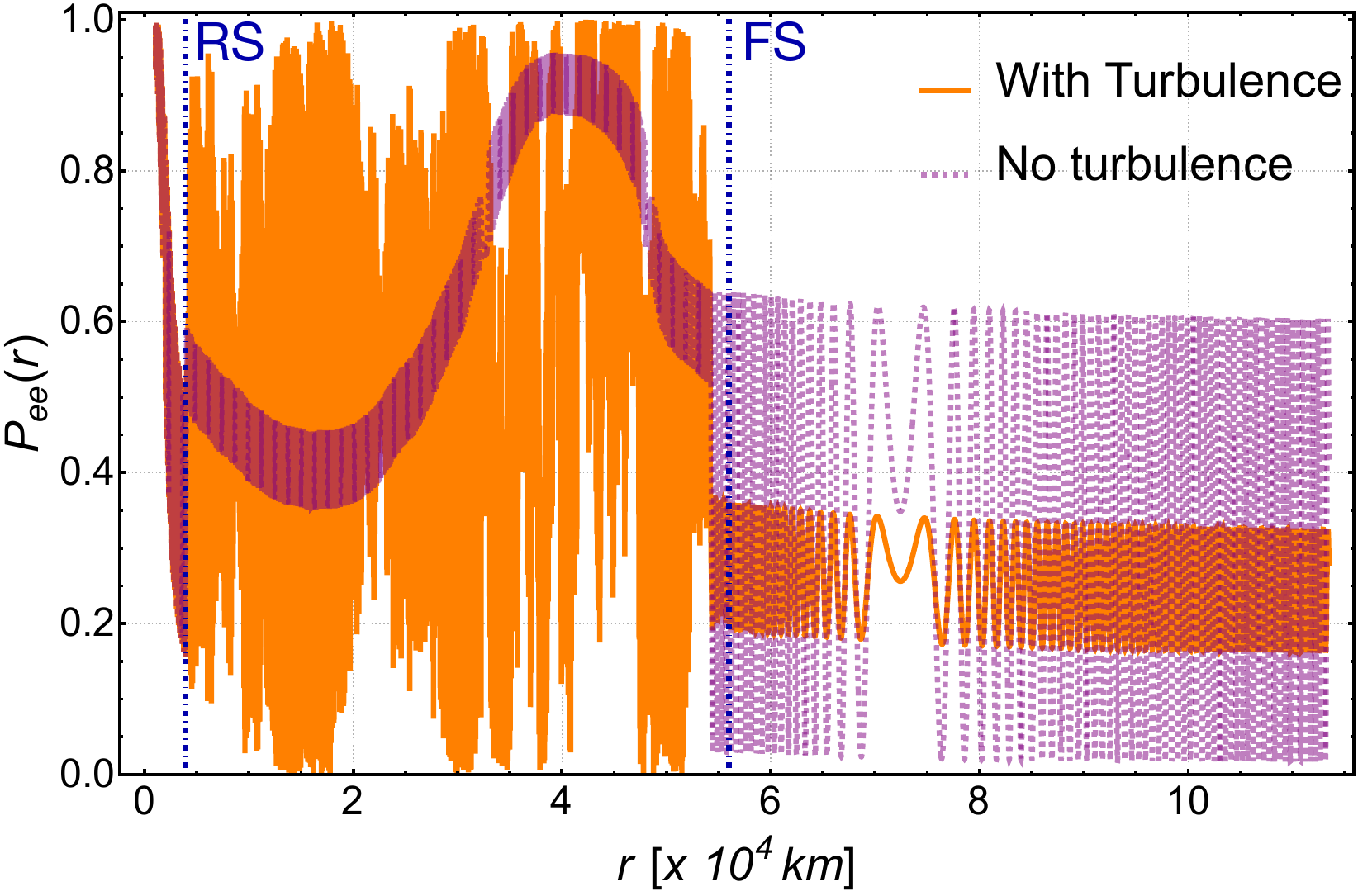}\label{fig:pee_comp}}\hfill
~~
\subfloat[(b)] {\includegraphics[width=0.49\textwidth]{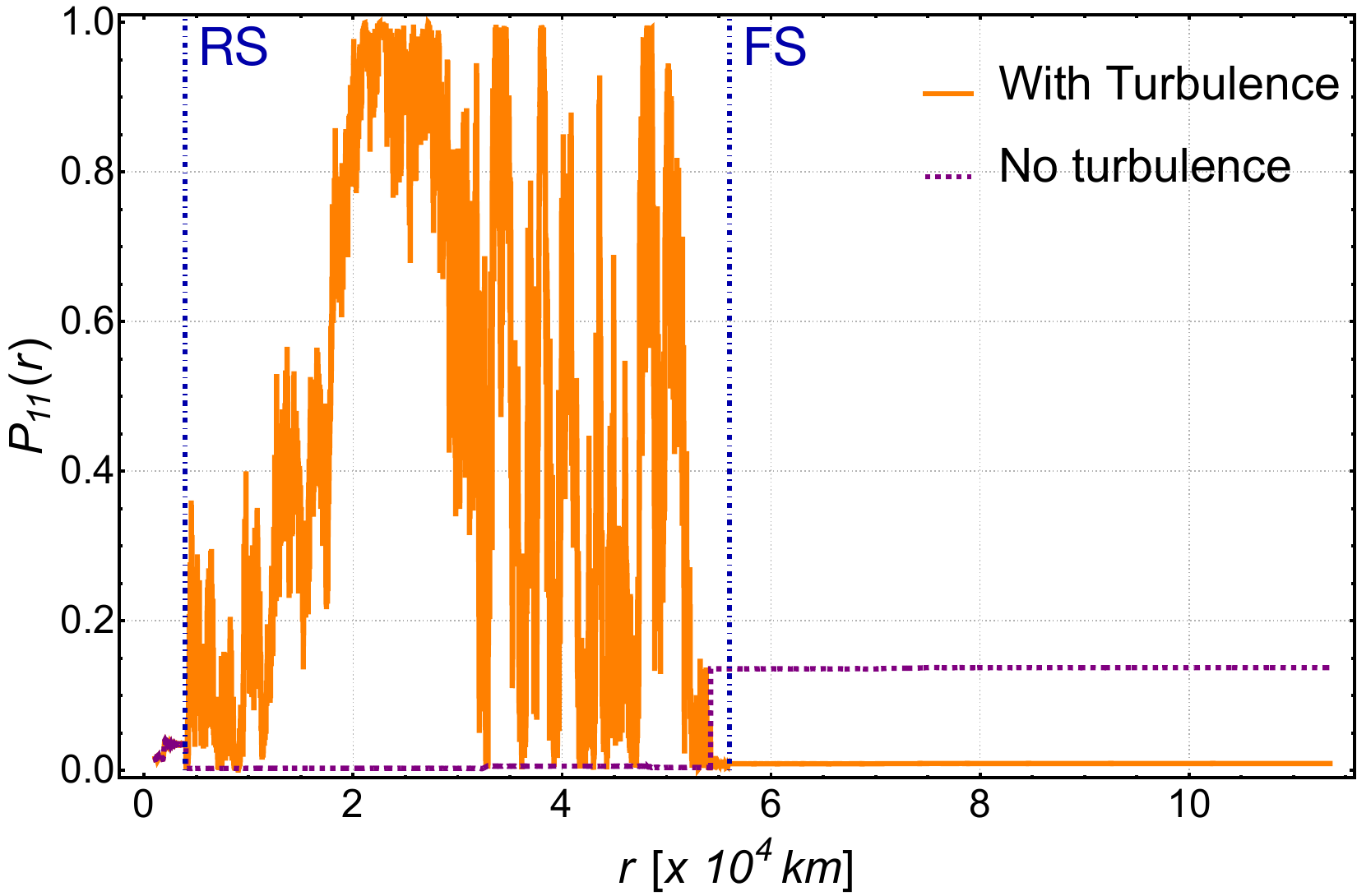}\label{fig:p11_comp}} \hfill
\\
\subfloat[(c)] {\includegraphics[width=0.49\textwidth]{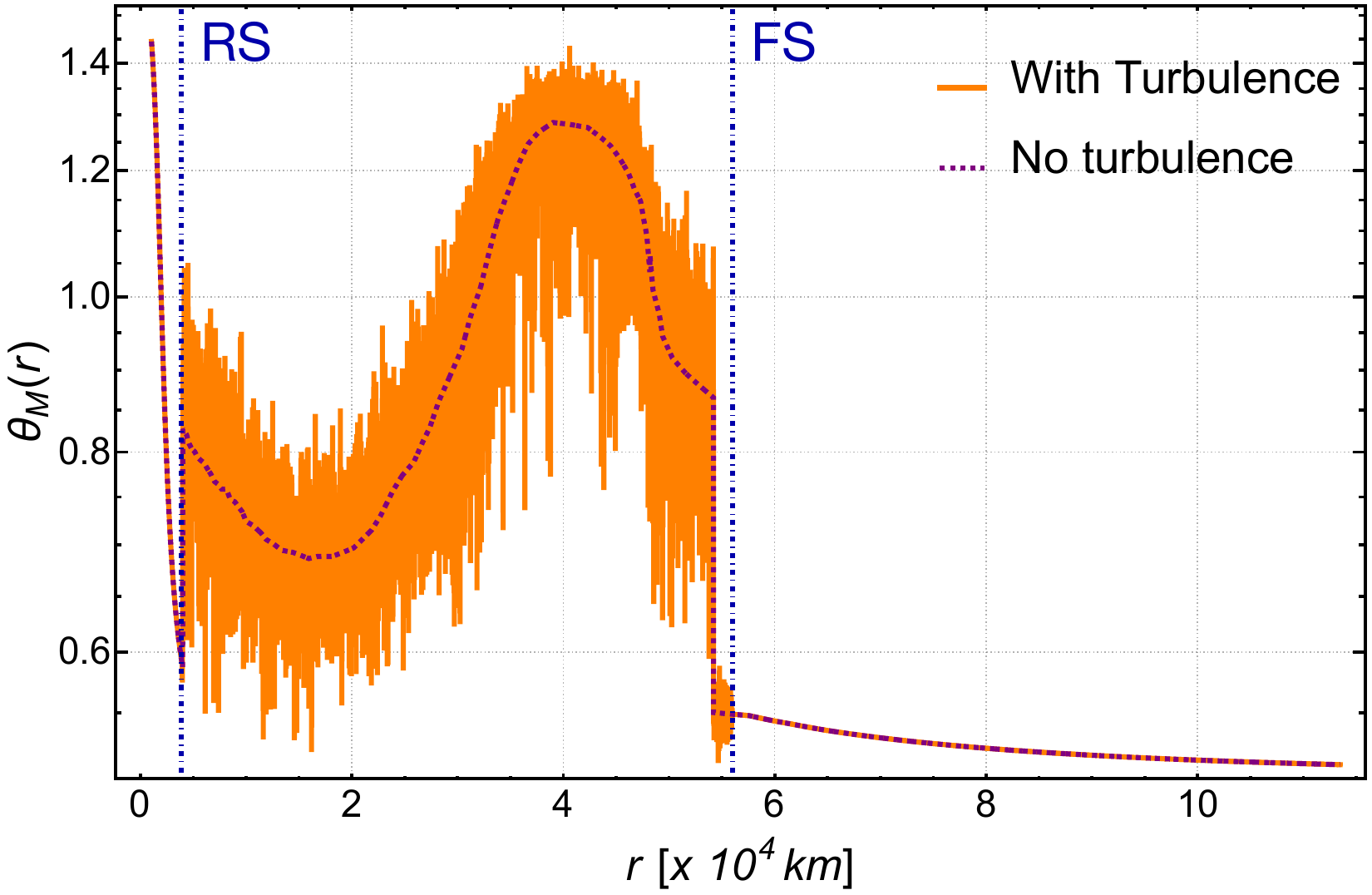}\label{fig:thetam_comp}} \hfill
~~
\subfloat[(d)] {\includegraphics[width=0.49\textwidth]{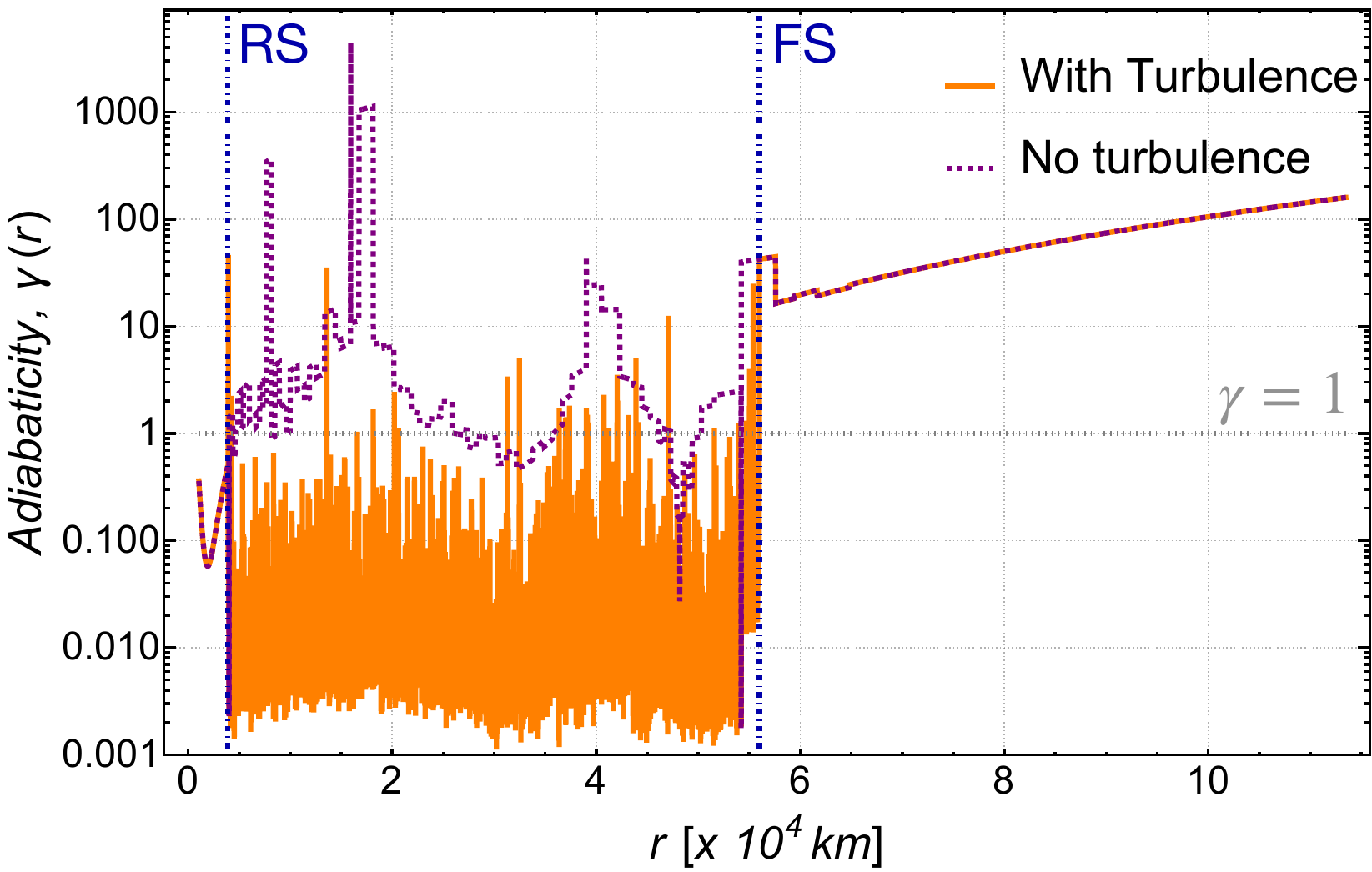}\label{fig:adia_comp}} \hfill
\caption{\label{fig:p_comp} Comparison between the scenarios when turbulence is present ($C_* = 0.3, \alpha = -5/3$) (solid orange curves) and when turbulence is absent ($C_* = 0$) (dashed purple curves). (a) The $\nu_e$ survival probability, $P_{ee}(r)$. (b) The survival probability for the first instantaneous matter eigenstate, $P_{11}(r)$. (c) The effective mixing angle in matter, $\theta_M(r)$ from Eq.~\eqref{eq:thetam}; (d) The adiabaticity parameter, $\gamma(r)$, from Eq.~\eqref{eq:adia}. The reverse shock (RS) and forward shock (FS) are shown as dot-dashed dark blue lines. We assume $\theta_0 = 0.09967$ rads, $\Delta m_{vac}^2 = 2.6 \times 10^{-3} \rm eV^2$, and $E_\nu = 10$ MeV.
}
\end{figure}
In this section, we discuss the results obtained by solving for neutrino flavour propagation through the matter profile as shown in Fig.~\ref{fig:turbprofs}. All the results shown correspond to the normal mass ordering of neutrinos unless otherwise stated. The two primary scenarios we are interested in are when turbulence is absent and when turbulence has a high amplitude. Thus, we compare the results for $C_* = 0$ (which we label as \emph{without turbulence}) and $C_* = 0.3$ (which we label as \emph{with turbulence})  respectively to implement the above scenarios. The choice of $C_* = 0.3$ is representative of an extreme case where complete flavour depolarization occurs and the oscillation signatures are washed out.

The results are shown in Fig.~\ref{fig:p_comp} for the electron neutrino survival probability ($P_{ee}(r)$), the survival probability for the first instantaneous matter eigenstate ($P_{11}(r)$), the effective mixing angle in matter ($\theta_M$) and the adiabaticity parameter ($\gamma(r)$) as $\nu_e$ propagates through the effective matter profile from $r_{\rm start}$ to $r_{\rm end}$. 
In Fig.~\ref{fig:pee_comp} we show the survival probability ($P_{ee} (r)$) for $\nu_e$ as a function of distance for the cases of with (solid orange) and without (dotted purple) turbulence. We note that the survival probability initially starts from unity, corresponding to a pure $\nu_e$ state. The position of reverse shock introduces a drop in the survival probability. Between the forward and the reverse shocks, the behaviour of $P_{ee}(r)$ is clearly distinct for the cases with and without turbulence. The propagation of the neutrinos in the presence of turbulence is extremely non-adiabatic due to the highly fluctuating background matter density. A better understanding can be achieved by studying $P_{11}(r)$ (in Fig.~\ref{fig:p11_comp}), which is expected to track the changes in the density. The matter eigenstate survival probability shows a jump due to non-adiabatic evolution corresponding to the density jumps. While the case without turbulence has jumps only at the positions of the reverse and forward shocks, the case with turbulence has numerous jumps. 

A similar trend can be seen in the effective mixing angle in the matter for the two cases, as shown in Fig.~\ref{fig:thetam_comp}. The case without turbulence exhibits a maximal change in $\theta_M$ at the locations of the forward and reverse shocks, whereas, in the presence of turbulence, $\theta_M$ has large jumps throughout the density profile. To demonstrate the non-adiabaticity associated with the neutrino flavour evolution, Fig.~\ref{fig:adia_comp} shows the associated adiabaticity parameter, $\gamma (r)$. A highly non-adiabatic evolution is characterized by $\gamma(r) << 1$. We see that for the case with turbulence between the reverse and the forward shocks, the adiabaticity parameter is much smaller than 1, implying a highly non-adiabatic evolution. The turbulence is only seeded between the forward and reverse shocks, so beyond the forward shock, the evolution once again becomes adiabatic in the absence of any further shocks or turbulence. This can be seen from Figs.~\ref{fig:p11_comp} and~\ref{fig:adia_comp}, where in the former case,  the transition probability does not show any jumps beyond the forward shock, whereas in the latter case, $\gamma(r) >> 1$ in the same region, implying adiabatic evolution.
\subsection{On the double-dip associated with shocks}
\label{sec:dbldip}
%%%%%%%%%%  %%%%%%%%%%%
%
The presence of shocks characterized by rapid changes in the density profile can have imprints on the neutrino oscillation probabilities. The MSW resonance associated with neutrino oscillations is given by, $\Delta m^2 {\rm cos} 2 \theta/ (2 E_\nu)  = \pm \sqrt{2} G_F Y_e \rho$, where $G_F$ is the Fermi constant, $Y_e$ is the electron fraction, and $\rho$ is the matter density. Depending on the mass-squared difference corresponding to solar or atmospheric neutrinos, the resonances are classified as L-resonance (low density) and H-resonance (high density) respectively. The propagation of the shockwave through the densities corresponding to H-resonance regions can break adiabaticity leading to distinctive features in the neutrino energy spectra. 

%
%%%%%%%%%% Discuss double-dip %%%%%%%%%%%
%
As discussed in Sec.~\ref{sec:intro}, numerical simulations hint towards the presence of both forward and reverse shocks in the matter density profile, resulting in a \emph{double-dip} feature in the neutrino spectra~\cite{Tomas:2004gr}. 
%
%%%%%%%%%% Discuss results with no turbulence%%%%%%%%%%%
%
%
\begin{figure}[!t]
\centering
\includegraphics[width=0.75\textwidth]{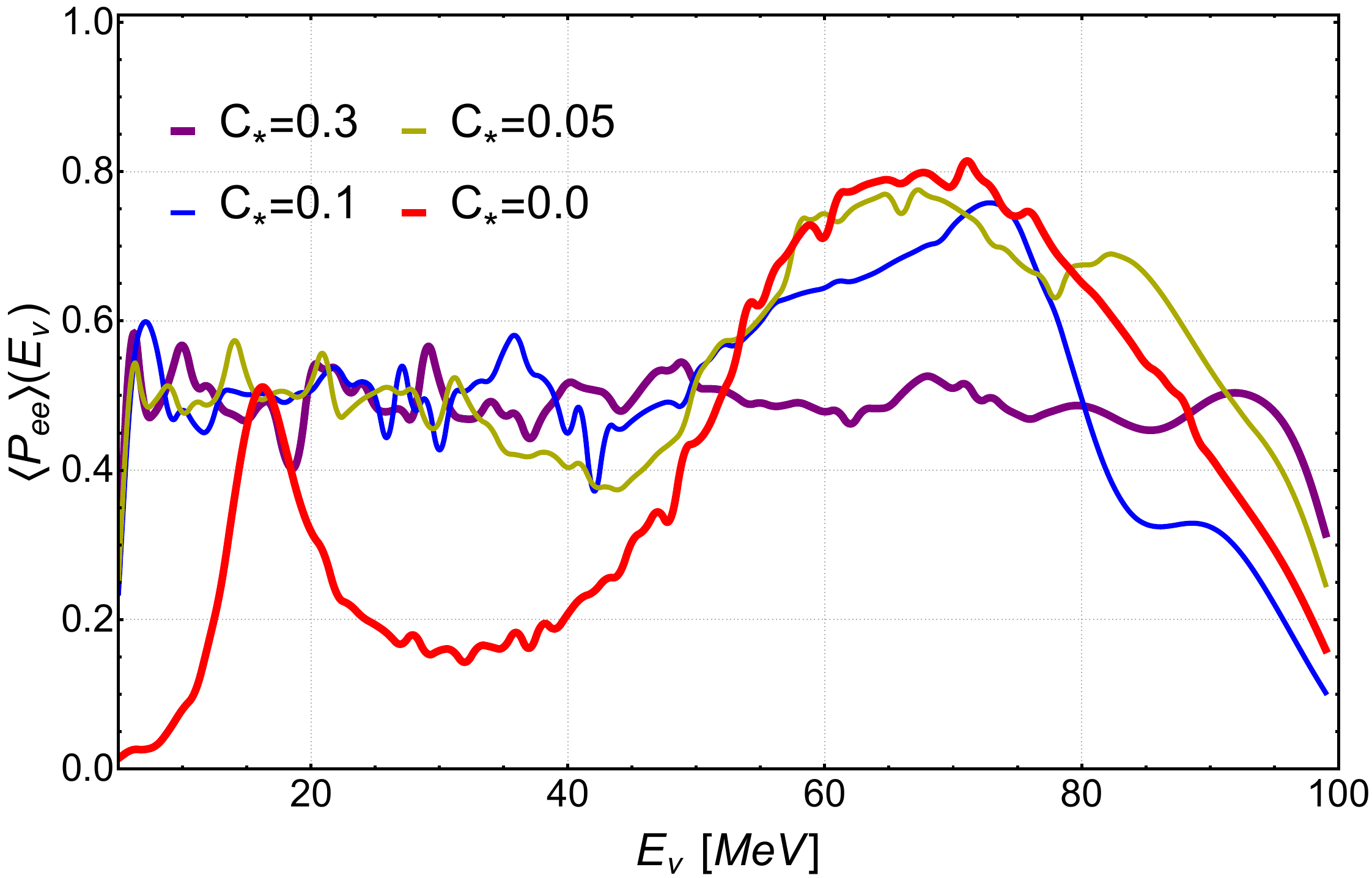}
\caption{\label{fig:dbl_dip_turb} The average survival probability ($\langle P_{ee} \rangle$) with neutrino energy ($E_\nu$) for different amplitudes of turbulence associated with the density profile (see Fig.~\ref{fig:turbprofs}).
}
\end{figure}
In Fig.~\ref{fig:dbl_dip_turb} we show the average survival probability for $\nu_e$ as a function of the neutrino energy ($E_\nu$). Note the presence of the double-dip feature (solid red curve) in the neutrino spectra corresponding to the density profile (in Fig.~\ref{fig:turbprofs}) in the absence of turbulence $(C_* = 0)$. The first peak corresponds to $E_\nu \sim 18$ MeV and is due to the forward shock. Following the first peak, the adiabatic transitions corresponding to the forward and reverse shocks partially cancel out to give the valley. Finally, a second peak appears for $E_\nu \gtrsim 60$ MeV which is due to the neutrinos crossing one of the forward or reverse shocks. We have checked that the double-dip feature also arises for individual realizations and is not an artefact of averaging the probability out.

%
%%%%%%%%%% Discuss results with turbulence%%%%%%%%%%%
%
We now aim to study the effects of the presence of turbulence on the double-dip feature in the neutrino spectra. The results are shown in Fig.~\ref{fig:dbl_dip_turb} for a range of turbulence amplitudes given a fixed value of $\alpha = -5/3$. The average survival probability $\langle P_{ee} \rangle$ is computed using a Gaussian energy resolution function, that is, 
\be
\langle P_{ee} \rangle (E_t) = \int_{E^{\rm lower}_r}^{E^{\rm upper}_r} \ dE_r P_{ee}(E_r) W(E_r,E_t)\,,
\ee
where, $W$ is the Gaussian energy resolution which depends on the intrinsic properties of the detector. The experimentally reconstructed energy in the detector is given by $E_r$ and the true neutrino energy is given by $E_t\equiv E_\nu$. The energy resolution function is defined as~\cite{Ohlsson:2000mj}
\be
\label{eq:energy_resfunc}
W (E_r,E_t) = \frac{1}{\sqrt{2 \pi} \sigma_E(E_r)} \exp{\bigg( - \frac{1}{2} \frac{(E_r - E_t)^2}{\sigma_E (E_r)^2} \bigg)}\,.
\ee
Interestingly, we find that the double-dip feature vanishes as the amplitude of turbulence increases leading to highly non-adiabatic flavour evolution. For $C_* > 0.01$, the first peak disappears. As the value of $C_*$ is increased, a gradual wash-out of the feature is seen. In fact, for $C_* \geq 0.3$, the survival probability averages to $1/2$, signifying complete flavour depolarization. Thus, the presence of turbulence can completely wash away the double-dip feature in the neutrino energy spectra associated with forward and reverse shocks. This can be used to probe the presence of hydrodynamic turbulence in core-collapse supernovae.  
%
%%%%%%%%%%%%%%%%%%
\subsection{Revisiting previous analytical results for a simplified model of turbulence}
%%%%%%%%%%%%%%%%%%%
In~\cite{friedland2006neutrino}, the effects of a simple model for turbulence on neutrino propagation in supernovae were studied. In particular, analytical estimates corresponding to neutrino flavour depolarization in the presence of a Kolmogorov-type turbulence were derived. The authors argued that evidence from numerical simulations hints towards the development of Kolmogorov cascades due to various instabilities. In such a turbulence scenario, the Fourier transform of the velocity correlator has the following form,
\be
C(k) \equiv \int dx \langle \delta n (0) \delta n(x) \rangle e^{-ikx} = C_0 k^\alpha\,,
\ee
where $\alpha = - 5/3$ and $C_0$ is a normalization associated with the density spectrum. The matter potential, $A(x)$, was assumed to be linear and turbulent random noise was introduced as
\be
A (x) = x + F \sum_{k=1}^{600} k^\beta \cos{\Big( kx + \phi (k) \Big)}\,,
\ee
where $F$ is the amplitude or normalization factor for the noise, $k$ is the mode number, $\beta = \alpha/2$, and $\phi(k)$ is a random phase which is a random value between $0$ and $2 \pi$ for every mode $k$. 

In this section, we follow the notations used in \cite{friedland2006neutrino} to consistently compare with their results. The amplitude $F$ used in the above equation is similar to $C_*$ in the prior sections, while $A(x)$ is a realisation of the charged current potential $V$ in Eq.~\eqref{eq:effpot}. Note that $C(k)$ defines the Fourier transform of the velocity correlator, whereas $E(k)$ in Eq.\,(\ref{eq:EK}) is the power spectrum associated with our method of seeding in the turbulence.

\begin{figure*}
\centering
\subfloat[(a)] {\includegraphics[width=0.5\textwidth]{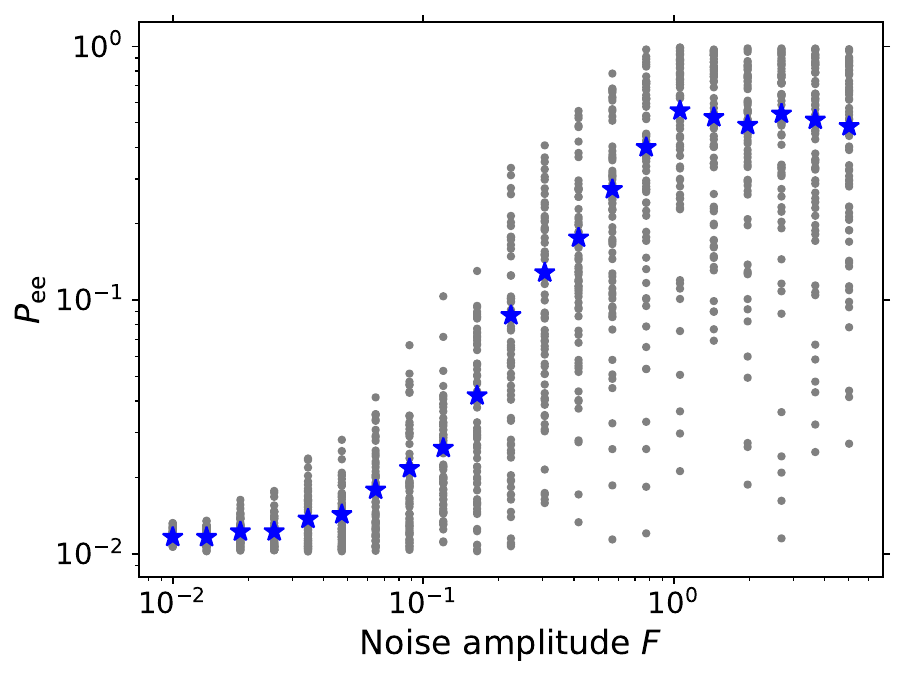}\label{fig:reprod}}\hfill
\subfloat[(b)] {\includegraphics[width=0.5\textwidth]{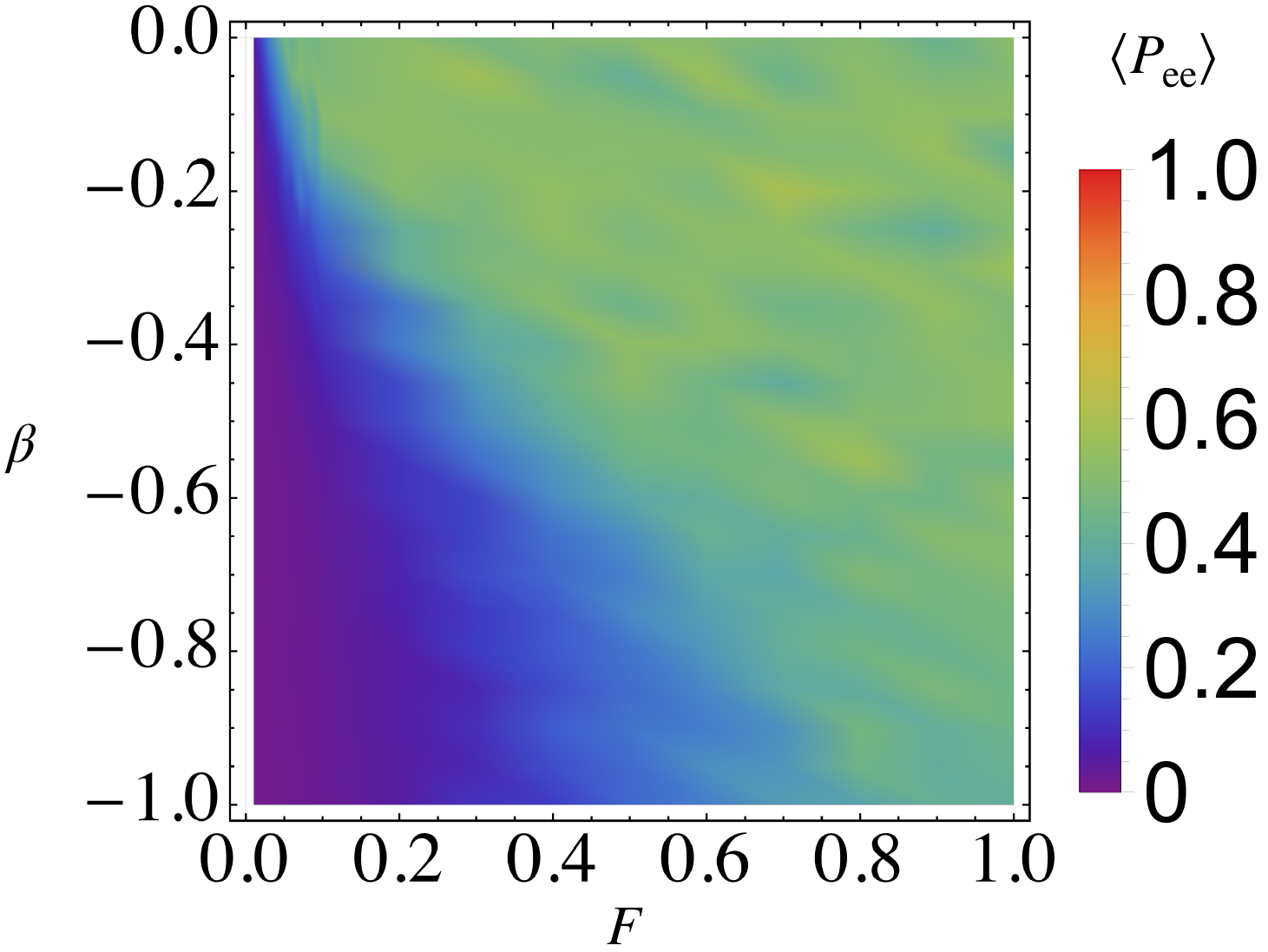}\label{fig:denplot}} \hfill
\caption{\label{fig:frdgruz1} (a) The electron neutrino survival probability as a function of the noise amplitude $F$ for $\beta = -5/6$, to match with the results of~\cite{friedland2006neutrino} (see Fig. 1 there).The grey dots represent the  $\nu_e$ survival probability ($P_{\rm ee}$) for each of the $66$ runs, while the blue stars represent the average $\nu_e$ survival probability ($\langle P_{\rm ee} \rangle$) over the $66$ runs. (b) The electron neutrino survival probability as a function of $\beta$ ranging from $-1$ to $0$ for various values of $F$.  Each data point in the numerics was repeated for $66$ times and the average value was used.
}
\end{figure*}

We reproduce the numerical results from~\cite{friedland2006neutrino} using the same parameters to check our numerical code. We show the result in Fig.~\ref{fig:reprod}. The spatial coordinate goes from $x=-100$ to $x = 100$, $\Delta = 50$, and $\theta = 0.1$. We choose $\Delta x = 0.1$. The agreement between our results and the same obtained by the previous work is good. We find that for low turbulence amplitudes, the effect on survival probability is less pronounced, which is expected and matches with our own analysis discussed above. When the turbulence amplitude is increased from $F \sim 0.2$ to higher values, a power law develops. The power law rise seen in the figure can be explained using an analytical estimate derived in~\cite{friedland2006neutrino},
\be
P^{\rm rise} \approx 0.84\ \frac{G_F C_0 (2 \Delta {\rm sin}\ 2 \theta_{13})^{-2/3}}{\sqrt{2} |n_0^\prime|}\,,
\ee
where $n_0^\prime$ is related to the spatial derivative of the matter density. We note that for values of $F \gtrsim 0.05$, a complete \emph{flavour depolarization} happens, implying the flavour constitution becomes equal for $\nu_e$ and $\nu_x$, in the two-flavour oscillation scenario considered here.

We extend the analysis in~\cite{friedland2006neutrino} by studying the variation of the average electron neutrino survival probability ($\langle P_{ee} \rangle$) in the $\beta - F$ plane. The result, in the form of a density plot for the average probability\footnote{The averaging is done over $66$ samples similar to Fig.~\ref{fig:reprod}.}, is shown in Fig.~\ref{fig:denplot}. A rainbow color scheme is used to illustrate $\langle P_{\rm ee} \rangle$, where the violet end of the spectrum implies $\langle P_{\rm ee} \rangle \sim 0$ and the red end signifies $\langle P_{\rm ee} \rangle \sim 1$. The area where complete flavour depolarization occurs is shown by a greenish-yellow shade corresponding to $\langle P_{\rm ee} \rangle \sim 0.5$. For $\beta \sim 0$, we find that $\langle P_{\rm ee} \rangle \sim 0.5$ for any given value of the turbulence amplitude $F$. As $\beta$ decreases to $-1$, higher amplitudes of turbulence lead to flavour depolarization. This generalizes the results of ~\cite{friedland2006neutrino} and confirms that for large values of $F$, flavour depolarization holds for other values of $\beta$ as well.

There are distinct differences between the analysis in~\cite{friedland2006neutrino} and our current work. We assume a more realistic matter density profile based on actual simulations in contrast to the linear profile assumed for simplicity in the former work. While the simple profile helps in getting some analytical estimates, the effects of a realistic profile need to be addressed numerically and lead to more realistic predictions for the upcoming detectors. Furthermore, the seeding of the turbulence in this work uses a variant of the Randomization method which is different from the simple random noise turbulence used in the previous work. 
\label{subsec:frdlnd}
\section{Events rates and statistical analysis}
\label{sec:evrates}
In previous sections, we discussed the effects of turbulence on the neutrino signatures and survival probabilities. This section focuses on the prospects and the capabilities of the upcoming neutrino detectors to constrain the parameters associated with turbulence, in particular, the amplitude $(C_*)$ and the spectral index $(\alpha)$. The detectors we consider for this work are the Deep Underground Neutrino Experiment (DUNE) and Hyper-Kamiokande (HK). While the former is largely sensitive to a flux of $\nu_e$, the latter is better suited for detecting the $\bar{\nu}_e$ flux.

The supernova neutrino spectrum at the source (the neutrinosphere) is assumed to be a pinched Fermi-Dirac spectrum~\cite{Keil:2002in},
\be
\frac{d N_\nu^{\rm src} (E_\nu)}{ d E_\nu}  = \frac{L_\nu}{\langle E _\nu \rangle^2} \frac{(1+\alpha_\nu)^{(1+\alpha_\nu)}}{\Gamma(1+\alpha_\nu)}\bigg( \frac{E_\nu}{\langle E_\nu \rangle} \bigg)^{\alpha_\nu} {\rm exp} \bigg( -(1+\alpha_\nu)\frac{E_\nu}{\langle E_\nu \rangle} \bigg)\,,
\ee
where $N_\nu^{\rm src}$ is the number of neutrinos at the source for any given flavour with energy $E_\nu$, $L_\nu$ is the neutrino luminosity for that particular flavour, $\langle E_\nu \rangle$ is the mean neutrino energy, $\alpha_\nu$ is a pinching parameter associated with the given flavour. Considering the cooling phase, we choose $L_{\nu_e} = L_{\bar{\nu}_e} =  5 \times 10^{51}$ erg, $L_{\nu_x} = L_{\bar{\nu}_x} =  10^{52}$ erg, $\langle E_{\nu_e} \rangle = 12$ MeV, $\langle E_{\bar{\nu}_e} \rangle = 15$ MeV, $\langle E_{\nu_x} \rangle = \langle E_{\bar{\nu}_x} \rangle = 18$ MeV, and $\alpha_{\nu_e} = \alpha_{\bar{\nu}_e} = \alpha_{\nu_x} = \alpha_{\bar{\nu}_x} = 3$. The neutrinos emitted from the neutrinosphere then propagate through the matter profile encountering the forward and reverse shocks, along with the turbulence which leads to neutrino oscillations. Let $P_{ee}(E_\nu)$ be the energy dependent $\nu_e$ survival  probability, obtained after solving for the propagation of the neutrinos through the SN envelope (realisations of these probabilities for different values of $C_*$ are shown in Fig.~\ref{fig:dbl_dip_turb}). In this case, the number of $\nu_e$ neutrinos at the end of the matter profile is given by
\be
\frac{d N_{\nu_e}^{\rm SN} (E_\nu)}{ d E_\nu} = P_{ee}(E_\nu) \frac{d N_{\nu_e}^{\rm src} (E_\nu)}{ d E_\nu} + \big( 1- P_{ee}(E_\nu) \big) \frac{d N_{\nu_x}^{\rm src} (E_\nu)}{ d E_\nu}\,,
\ee
where $N_{\nu_e}^{\rm SN}$ is the number of $\nu_e$ after the initial neutrino spectra from the neutrinosphere have propagated through the matter profile.
Similarly one can obtain the spectra for $\bar{\nu}_e$. Once the neutrinos leave the supernova they propagate through the interstellar medium to reach Earth. 
This can be effectively treated as propagation in vacuum\footnote{Note that by this time, the neutrinos will have decohered and will travel as an incoherent superposition of mass eigenstates.}. The vacuum survival probability for $\nu_e$ is given by 
\be
P_{ee}^{\rm vac} (E_\nu) = 1 - {\rm sin}^2 2 \theta\ {\rm sin}^2 \bigg( \frac{\Delta m_{\rm vac}^2 R}{4 E_\nu} \bigg) \simeq \cos^4\theta + \sin^4\theta = 0.98\,,
\ee
where $\theta_{\rm vac}$ is the vacuum mixing angle, $\Delta m^2_{\rm vac}$ is the vacuum mass-squared difference, and $R$ is the distance between the source and the Earth (for galactic scales this is fixed to $R = 10$ kpc). We assume $\theta_{\rm vac} = 0.09967$ rads and $\Delta m^2_{\rm vac} = 2.6 \times 10^{-3} \rm eV^2$. Thus the neutrino spectra at the Earth, neglecting Earth-matter effects, is given by
\be
\frac{d N_{\nu_e}^{\rm earth} (E_\nu)}{ d E_\nu} = P_{ee}^{\rm vac}(E_\nu) \frac{d N_{\nu_e}^{\rm SN} (E_\nu)}{ d E_\nu} + \big( 1- P_{ee}^{\rm vac}(E_\nu) \big) \frac{d N_{\nu_x}^{\rm SN} (E_\nu)}{ d E_\nu}\,.
\ee
A similar analysis can be repeated for obtaining the $\bar{\nu}_e$ spectra on Earth which will be relevant for HK. Note that, for our analysis, we consider $\nu_e$ in normal mass-ordering in DUNE and $\bar{\nu}_e$ in inverted mass-ordering in HK, as these are the channels sensitive to MSW resonances. Also, we do not consider Earth matter effects for this work.

The total number of neutrinos of a given species $\alpha$ ($N_{\nu_\alpha}$) detected in a neutrino detector from a source at a distance $R$ is given by
\be
N_{\nu_\alpha} = \Delta t \frac{N_{\rm tar}}{4 \pi R^2} \int d E_r \int d E_t\ \frac{d N_{\nu_\alpha}^{\rm earth}(E_t)}{d E_t} \sigma_{\nu_\alpha}(E_t) W (E_r,E_t)\,,
\ee
where $N_{\rm tar}$ is the number of targets for a given channel in the detector fiducial volume, $\sigma_{\nu_\alpha}$ is the neutrino interaction cross-section with the targets for the relevant channel, $\Delta t$ is the duration for which we assume the density profile to remain approximately fixed and similar to what is considered for the matter profile, and $W$ is the Gaussian energy resolution function (see Eq.~\eqref{eq:energy_resfunc}). We fix the distance to the supernova at $R = 10$ kpc for our current work. The sensitivity of the neutrino detectors for different values of $C_*$ and $\alpha$ is estimated using a simple $\chi^2$ estimator
\be
\label{eq:chisq}
\chi^2 = \frac{\big(N_{\rm obs} - N_{\rm exp} \big)^2}{N_{\rm obs}}\,,
\ee
where $N_{\rm obs}$ is the number of neutrino events observed in the neutrino detector, $N_{\rm exp}$ is the expected number of neutrino events in the detector. We fix $N_{\rm exp}$ to the case when $C_* = 0.1$, and $\alpha = 1.65$, corresponding to the Kolmogorov turbulence, unless otherwise specified. We consider $\Delta t=1\,{\rm s}$.
\begin{figure}[!t]
\centering
\includegraphics[width=\textwidth]{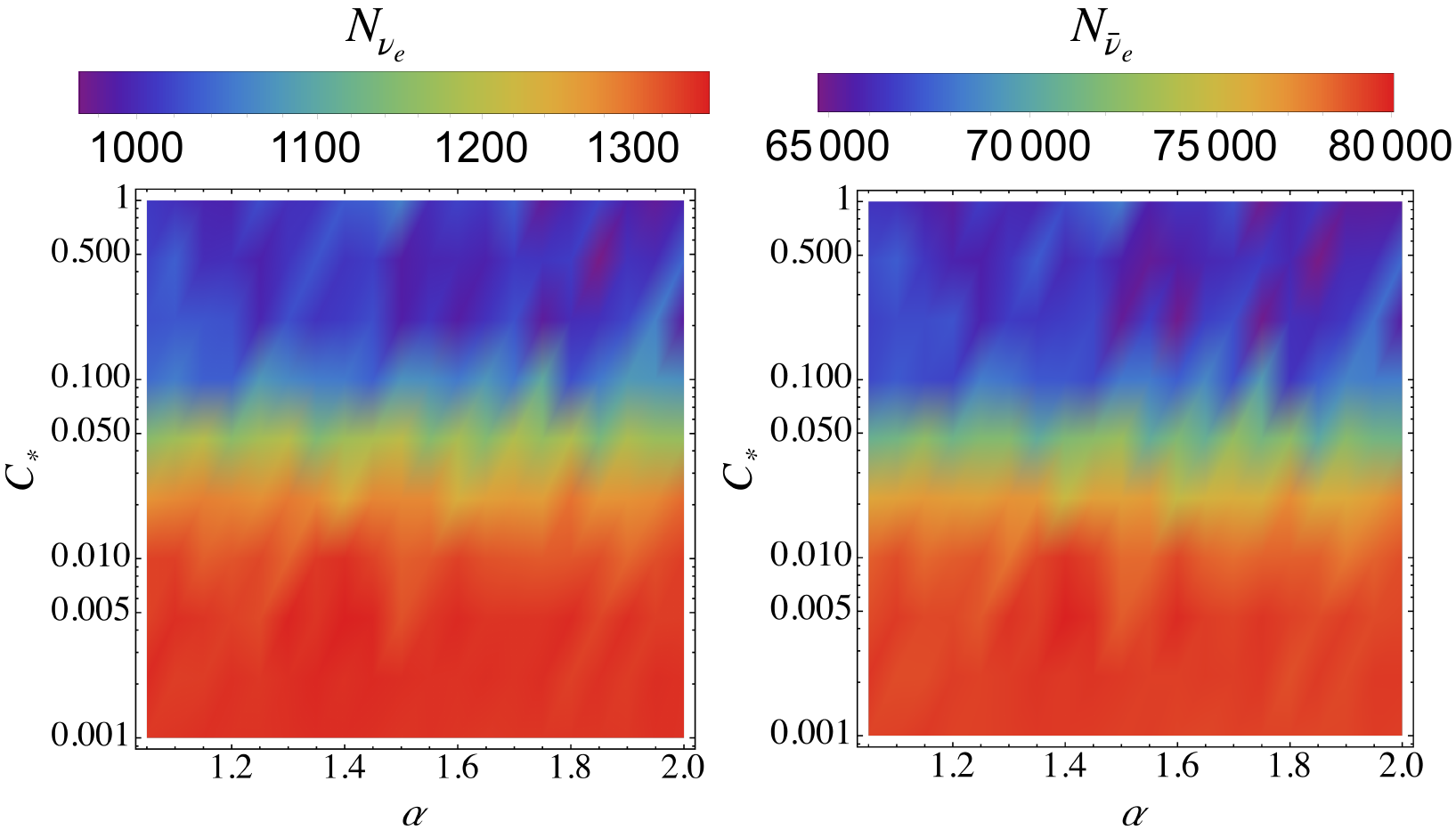}
\caption{\label{fig:cstar_alpha_scan} Density plot to show the total number of $\nu_e$ events for DUNE, assuming normal mass ordering (\emph{left}) and $\bar{\nu}_e$ events for HK, assuming inverted mass ordering (\emph{right}) on the $C_* - \alpha$ plane. The source is assumed to be at a distance of $R = 10$ kpc. See text for details on the other parameters used.
}
\end{figure}
\subsection{DUNE}
\label{subsec:dune}
The DUNE experiment~\cite{DUNE:2020lwj} consists of a near and a far detector separated by $1300$ km. The former is located at Fermilab, Illinois while the latter is located about $1.5$ km below the ground at the Stanford Underground Research Facility (SURF), South Dakota. The far detector is a liquid argon time projection chamber (LArTPC) with $70$ kt (fiducial mass of $40$ kt) of liquid argon. We will focus on the far detector in this work.

The relevant interaction channel for supernova neutrinos is the charged current interaction of electron neutrinos with argon nuclei, $\nu_e + ^{40} {\rm Ar} \rightarrow ^{40} {\rm K}^* + e^-$. We use $N_{\rm tar} = 6 \times 10^{32}$. The interaction cross-section $\sigma_{\nu_e}$ associated with the $\nu_e - \rm Ar$ interaction cross section is taken from~\cite{Jana:2022tsa}. The energy resolution for DUNE is taken to be $\sigma_E/ {\rm MeV} = 0.11 \sqrt{E_r/ {\rm MeV}} + 0.02 E_r/{\rm MeV}$.

The result for the total number of electron neutrino events $N_{\nu_e}$ for DUNE is shown in the left panel of Fig.~\ref{fig:cstar_alpha_scan} as a density plot in the $C_* - \alpha$ plane. We assume normal mass ordering. The color map uses a rainbow template to denote the lowest number of events in violet and the highest in red. For a typical galactic scale supernova ($R \sim 10$ kpc) we have $\sim 10^3$ events in DUNE in the $\nu_e$ channel. There are two interesting aspects that we note right away - (a) as $\alpha$ varies from $1.0$ to $2.0$, the number of events $N_\nu$ does not change significantly, which implies that it would be difficult to put constraints on the value of $\alpha$ using the DUNE detector; (b) as $C_*$ is varied from $0.001$ to $1.0$ the number of events in the DUNE detector has a significant variation and goes from $\sim 1350$ (red) for lower values of $C_* \sim 0.001$ to $\sim 900$ (violet) for large values of $C_* \sim 1.0$. Thus, DUNE can be used to place constraints on the value of $C_*$ for a typical galactic supernova in the near future. We discuss the two points below using a $\chi^2$ plot as discussed below.

\begin{figure}[!t]
\centering
\subfloat[(a)] {\includegraphics[width=0.5\textwidth]{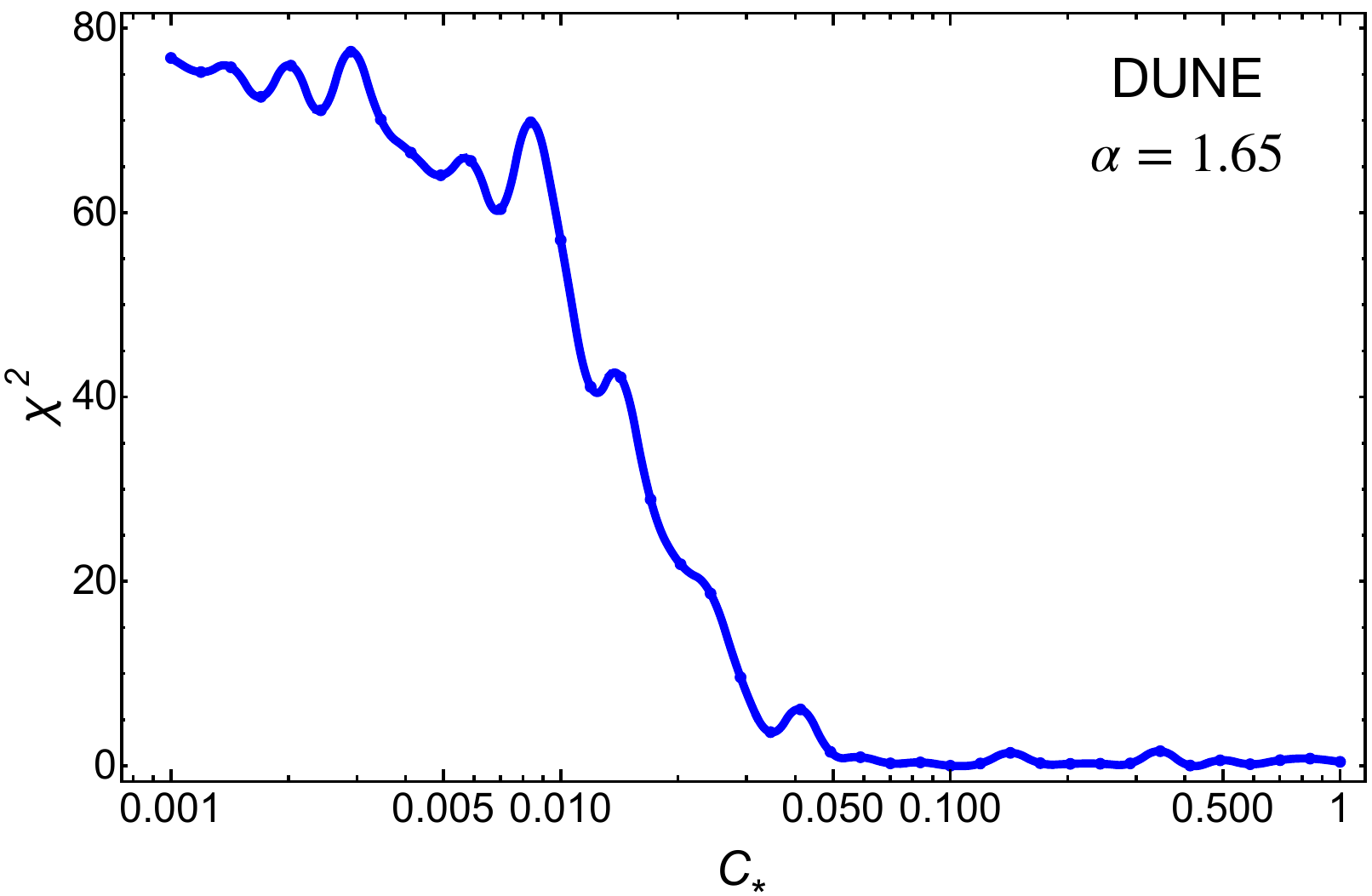}\label{fig:alphaComp_dune}}\hfill
\subfloat[(b)] {\includegraphics[width=0.5\textwidth]{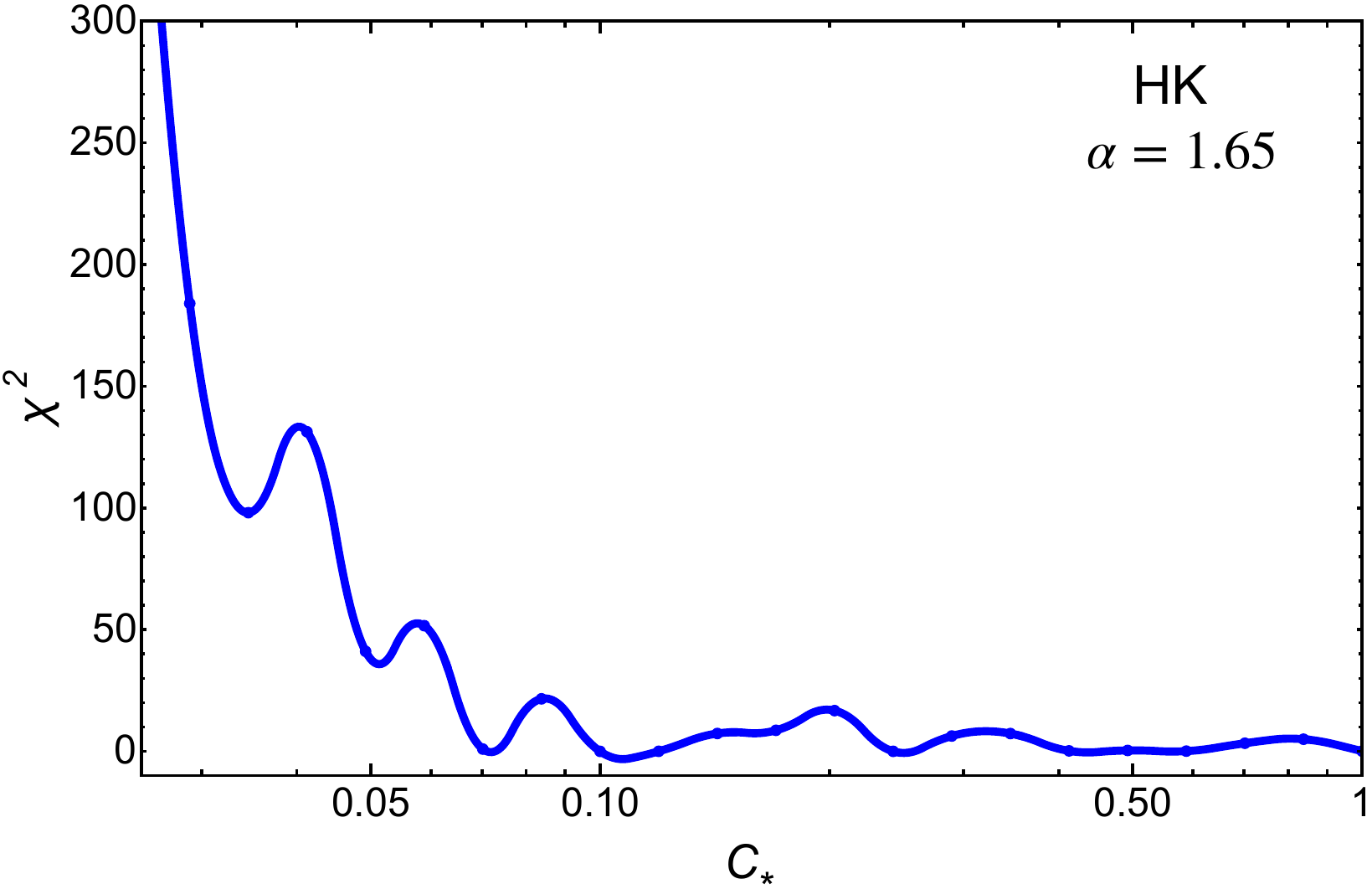}\label{fig:alphaComp_hk}} \hfill \\
\subfloat[(c)] {\includegraphics[width=0.5\textwidth]{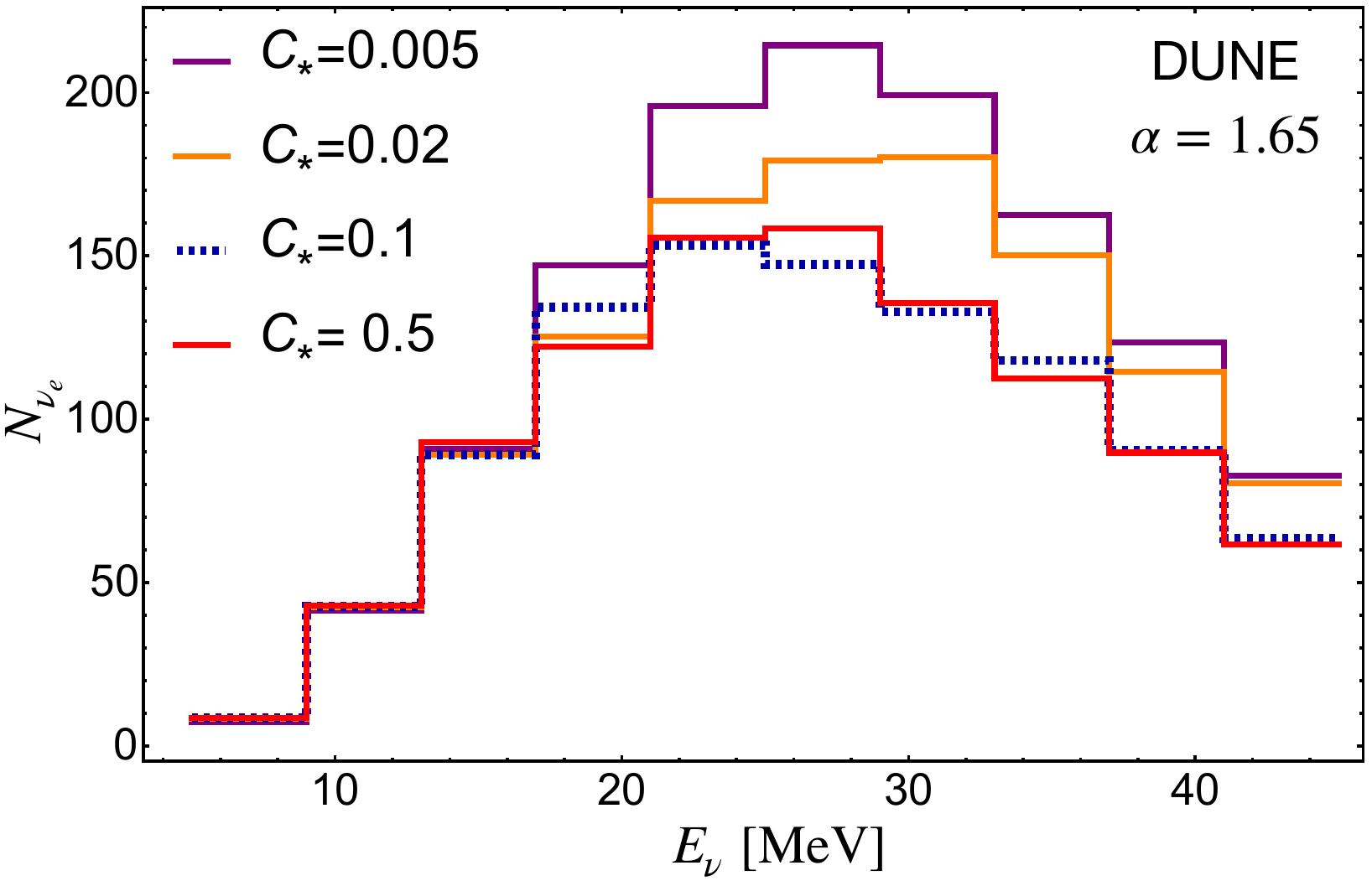}\label{fig:alphaComp_casehiXi_dune}}\hfill
\subfloat[(d)] {\includegraphics[width=0.5\textwidth]{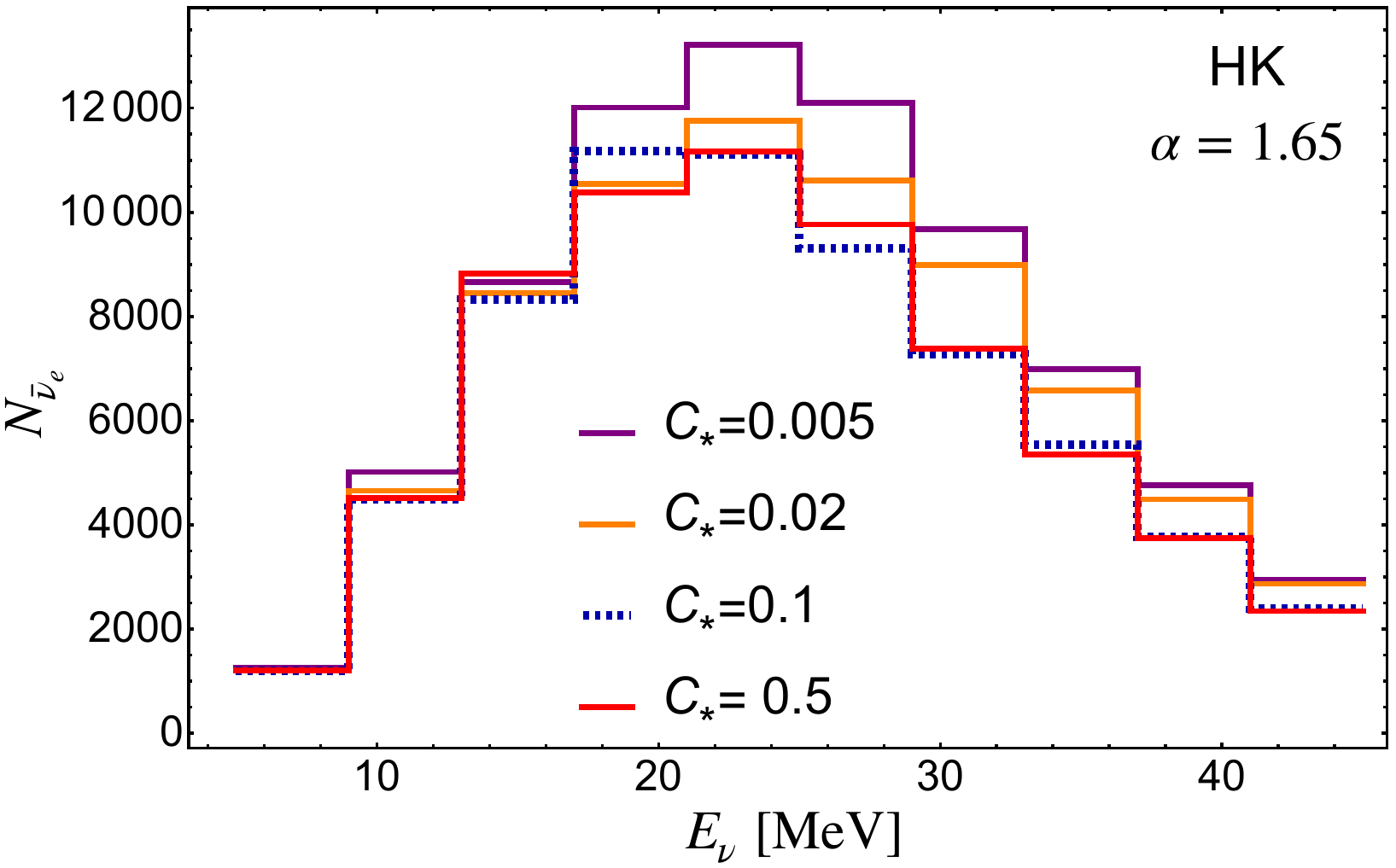}\label{fig:alphaComp_casehiXi_hk}} \hfill
\caption{\label{fig:var_cstar} (a) Plot of $\chi^2$ estimator for different values of $C_*$ for DUNE assuming $\alpha = 1.65$ (corresponding to Kolmogorov turbulence); (b) same as (a) but for HK; (c) The expected number of $\nu_e$ events, assuming normal mass ordering per energy bin for DUNE assuming $\alpha = 1.65$ and $C_* = 0.005$ (solid purple), $0.02$ (solid orange), and $0.5$ (solid red). The $N_{\rm exp}$ for the $\chi^2$ estimator is considered to be the case when $C_* = 0.1$ and $\alpha = 1.65$, which is shown as a dotted dark blue line; (d) same as (c) but for HK where $\bar{\nu}_e$ events, assuming inverted mass ordering are considered.
}
\end{figure}

The significant variation of the number of events with $C_*$ is confirmed in Fig.~\ref{fig:alphaComp_dune}, which shows the variation of $\chi^2$ versus $C_*$ for $\alpha = 1.65$. The null hypothesis is considered to be $C_* =0.1$. We note that for low values of $C_* < 0.05$, the $\chi^2$ starts growing and it increases as the amplitude of turbulence decreases\footnote{Note that for $C_* = 0.1$ the value of $\chi^2$ goes to $0$ from Eq.~\eqref{eq:chisq}.}. However, for larger values of $C_* > 0.05$, constraints become weak. To get a better understanding of this behaviour, we compare the event spectra for different values of $C_*$ in Fig.~\ref{fig:alphaComp_casehiXi_dune}. The null hypothesis $(C_* = 0.1)$ is shown with a dotted dark blue line. The cases corresponding to $C_* = 0.005, 0.02$, and $0.5$ are shown as solid purple, orange, and red lines. On comparing the case of $C_* = 0.005$ with the expected case we see a significant difference in the number of events in the energy bins above $17$ MeV leading to the high value of $\chi^2$. As $C_*$ increases the number of events in the energy bins becomes similar to the expected case of $C_* = 0.1$ explaining the trend of $\chi^2$ as seen in Fig.~\ref{fig:alphaComp_dune}.

\begin{figure*}
\centering
\subfloat[(a)] {\includegraphics[width=0.49\textwidth]{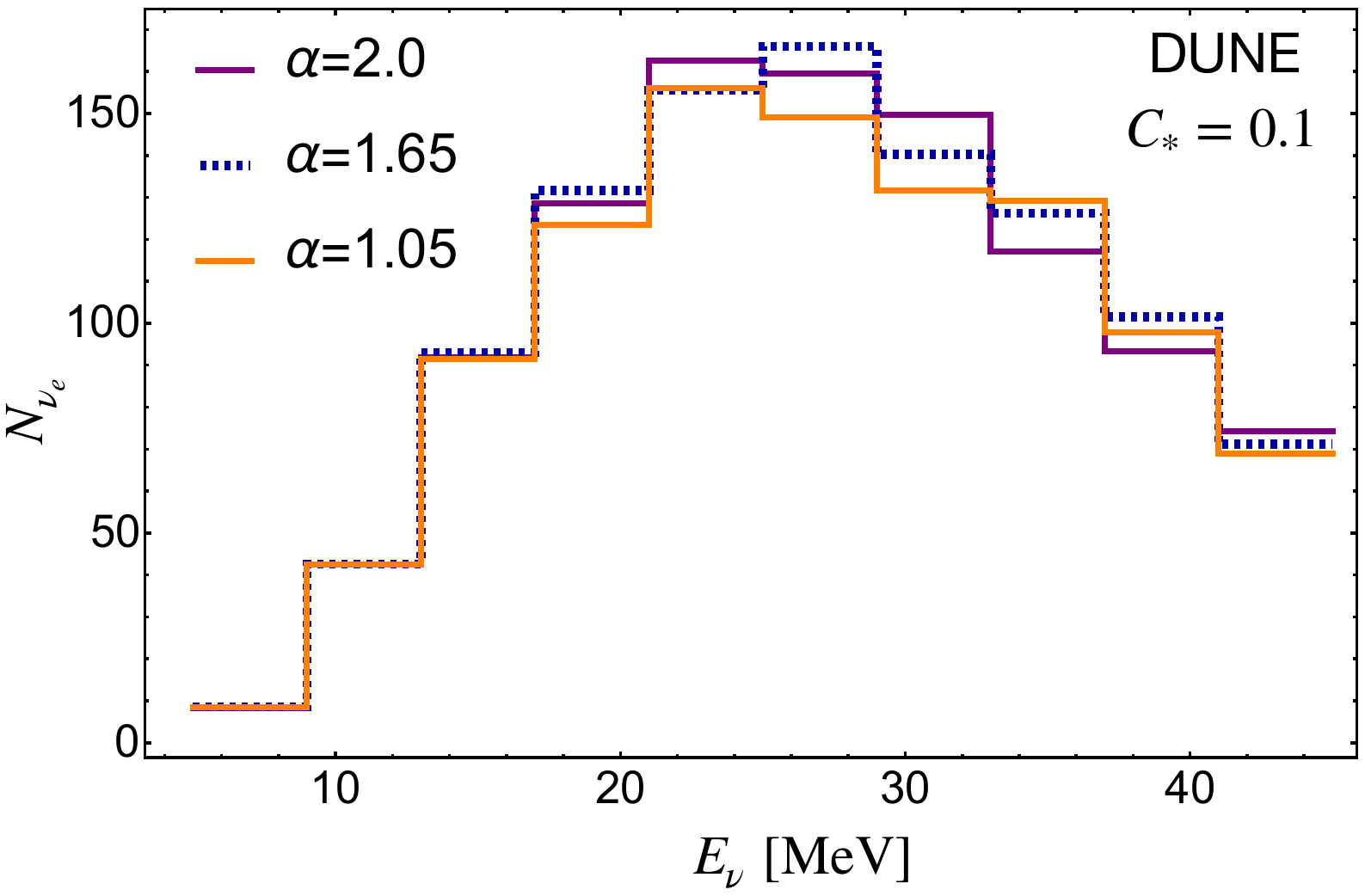}\label{fig:alphaComp_caselowXi_dune}}\hfill
\subfloat[(b)] {\includegraphics[width=0.49\textwidth]{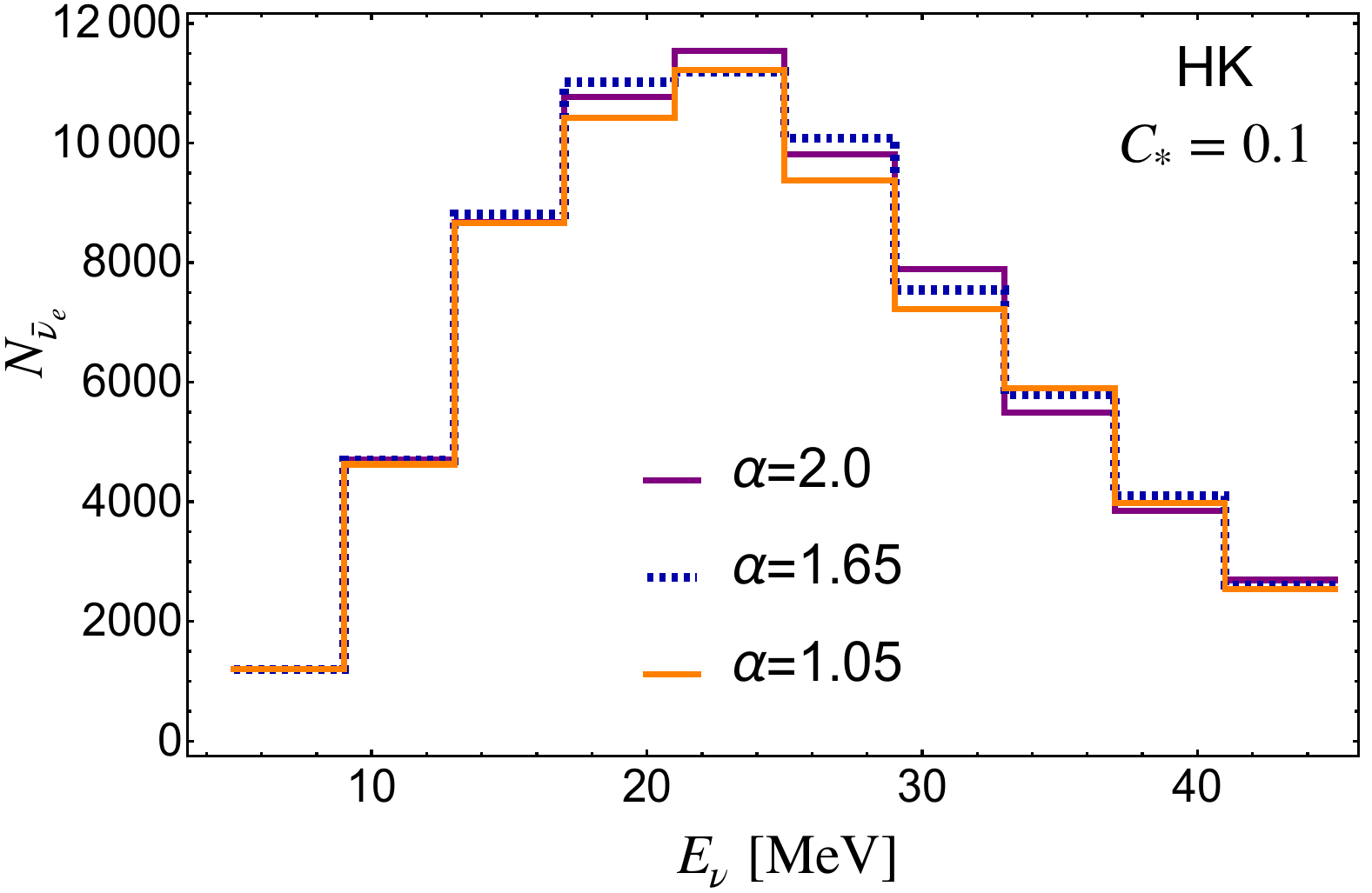}\label{fig:alphaComp_caselowXi_hk}} \hfill
\caption{\label{fig:var_alpha} The expected number of $\nu_e$ and $\bar{\nu}_e$ events per energy bin corresponding to DUNE (\emph{left}) and HK (\emph{right}) for a fixed value of $C_* = 0.1$ and different values of $\alpha = 1.05$ (solid orange), $1.65$ (dotted dark blue), and $2.0$ (solid purple) respectively. The source is assumed to be at a distance of $R= 10$ kpc.
}
\end{figure*}

The variation over $\alpha$ does not have a specific trend for any value of $C_*$ and the $\chi^2$ values obtained are extremely low and we do not show them here. This is evident from Fig.~\ref{fig:alphaComp_caselowXi_dune} where we show the event spectra for $\alpha = 1.05$ (solid orange), $1.65$ (dashed dark blue), and $2.0$ (solid purple) for $C_* = 0.1$. The expected scenario for this case is again chosen to be $C_* = 0.1$ and $\alpha = 1.65$ shown as the dashed dark blue line. We note that the spectra for all the three cases shown look very similar to the expected case in dotted dark blue, explaining the low values of $\chi^2$. This highlights the ineffectiveness of DUNE to put constraints on the value of $\alpha$. 
\subsection{Hyper-Kamiokande}
Hyper-Kamiokande (HK)~\cite{Hyper-Kamiokande:2022smq} is an upcoming $260$ kton (fiducial $187$ kton) water-Cherenkov detector, being built at the Kamioka Observatory in Japan. It is an upgrade to the Super-Kamiokande detector but with around $10$ times the fiducial volume. It will consist of two tanks of ultrapure water acting as the medium for neutrino detection. The primary detection channel in HK is the inverse-beta decay (IBD), $\bar{\nu}_e + p \rightarrow e^+ + n$. We use the IBD cross-section from~\cite{Strumia:2003zx}. Similar to Super-Kamiokande, the energy resolution for HK is given by $\sigma_E/ {\rm MeV} = 0.6 \sqrt{E_r/ {\rm MeV}}$~\cite{Jana:2022tsa}. The number of targets for HK is assumed to be $1.2 \times 10^{35}$.

The density plot for the number of $\bar{\nu}_e$ events for HK in the $C_* - \alpha$ plane is shown in the right panel of Fig.~\ref{fig:cstar_alpha_scan}. The details of the colors used are discussed in Sec.~\ref{subsec:dune}. We have $\sim 7 \times 10^4$ $\bar{\nu}_e$ events for a typical galactic scale supernova ($R = 10$ kpc). This number is much higher than DUNE owing to the larger number of targets. This already hints at the possibility that HK would be a better detector than DUNE at constraining the parameter range for turbulence. However, similar to DUNE, the variation in $N_{\bar{\nu}_e}$ is insignificant across the different values of $\alpha$ for a given value of $C_*$.

Using HK, relatively strong constraints can be placed on the amplitude of turbulence as seen by the variation in $N_{\bar{\nu}_e}$ with $C_*$ for a fixed value of $\alpha$ in the right panel of Fig.~\ref{fig:cstar_alpha_scan}. In this case, we assume inverted mass ordering. This is also evident from Fig.~\ref{fig:alphaComp_hk} where we have very large values of $\chi^2$ for $C_* \lesssim 0.05$, keeping $\alpha$ fixed at $1.65$. To confirm our results, we show the event spectra for $\alpha = 1.65$ and different values of $C_*$ in Fig.~\ref{fig:alphaComp_casehiXi_hk}. Once again, the expected case is assumed to be $C_* = 0.1$ and $\alpha = 1.65$ and is shown in dotted dark blue. We note that similar to the case of DUNE, for $E_\nu \leq 20$ MeV, the spectra are similar among all the cases, however for $E_\nu > 20$ MeV, there is a significant difference in the spectra between $C_* = 0.1$ and $C_* = 0.005$ (solid purple line) and $C_* = 0.02$ (solid orange line) leading to the comparatively high values of $\chi^2$. However, for $C_* = 0.5$ (solid red line) the event spectra look similar to the expected case giving a low value of $\chi^2$ as seen in Fig.~\ref{fig:alphaComp_hk}. 

Finally in Fig.~\ref{fig:alphaComp_caselowXi_hk}, we show the event spectra for $C_* = 0.1$ for different values of $\alpha$. The spectra look similar to the expected case of $C_* = 0.1$, indicating that HK, while sensitive, would not be able to put very strong constraints on $\alpha$.

\section{Conclusions and Implications}
\label{sec:concl}
Neutrino propagation through a turbulent medium leads to high degrees of non-adiabaticity. Hydrodynamical simulations of core-collapse supernova hint towards the presence of various instabilities in the region behind the shock, leading to the matter profile becoming turbulent. Such a turbulent medium can leave its signature on the neutrino spectra, thereby allowing future neutrino detectors to probe the strength of turbulence inside a core-collapse supernova.
In this work, we performed a detailed study of the effects of turbulence on neutrino propagation and outlined the prospects for upcoming neutrino detectors like DUNE and HK to constrain the parameter space associated with turbulence.

We considered a usual supernova matter profile snapshot, characteristic of the cooling phase, and used the variant C formalism of the Randomization method to seed in the turbulence. This involves a stratified sampling of the modes to have a minimal effect associated with the physical scales. The two main parameters associated with turbulence are the amplitude ($C_*$) and the spectral index associated with the power spectrum ($\alpha$). We numerically evolved the neutrino oscillation probabilities using the slab approximation, which is appropriate for a highly oscillatory profile as is the case in the presence of turbulence. It is important to note that for getting reasonable results, the discretization scale, which in this case is given by the width of the slabs, should be smaller than the length scales associated with the variation of the matter profile. To illustrate the difference between neutrino propagation in the presence and absence of turbulence, we contrasted various quantities like the electron neutrino survival probability, the effective mixing angle in matter, and the adiabaticity parameter. We showed that these quantities change significantly depending on whether the matter background has turbulence or not.

Depending on the location of the forward and reverse shocks, the neutrino spectra from a supernova can be very different due to the non-adiabaticity associated with the shock fronts. In particular, the H-resonance associated with $\Delta m^2_{\rm atm}$ can break adiabaticity at the forward and reverse shocks. This can lead to a double-dip structure in the neutrino survival probability, as was discussed in detail in~\cite{Tomas:2004gr}. We revisited this issue and studied if the double-dip feature still survives in the presence of turbulence between the shockfronts. We found that as the profile becomes more turbulent, flavour depolarization kicks in. In fact, in our study, if the strength of turbulence exceeds $30\%$, the double-dip structure is washed out, leading to complete flavour depolarization. This implies that the double-dip signature which can provide more information about the positions of the forward and reverse shocks may fail to do so in the presence of strong turbulence in between the shocks.

We considered the prospects of upcoming neutrino detectors like DUNE and HK to detect neutrinos from a galactic core-collapse supernova and constrain the parameters associated with turbulence in Sec.~\ref{sec:evrates}. Given the larger fiducial volume, HK is better in constraining the turbulence parameters than DUNE. Most importantly, we note that there is a significant variation of the number of events as the amplitude increases from sub-percent to $\mathcal{O}(1)$ for a fixed value of the spectral index. As a result, both DUNE and HK would be able to place reasonable constraints on the amplitude of the turbulence. However, the same cannot be said for the spectral index as the sensitivity is not too large. Note that this would be an indirect probe of the presence of turbulence inside the SN. The nature and strength of the turbulence can only be robustly determined through large-scale hydrodynamic simulations.

We conclude by discussing potential improvements in our results. With regard to the turbulence, we assumed a simplified model of one-dimensional to illustrate our results. A more realistic picture can be obtained using three-dimensional computations where the turbulence samples $k_x, k_y$, and $k_z$ from given power spectral distributions. However, this would be computationally intensive. Furthermore, we also assumed that the matter density profile along with the turbulence is constant for the duration of neutrino propagation we consider, which is set to $1$ second. This is a reasonable approximation since the density profiles from numerical simulations do not have significant changes over such durations in the cooling phase.
It is important to note however, that for subsonic shocks the variation in the matter density profile is small, which implies $C_* << 0.1$. This leads to our results being optimistic to constrain lower amplitudes of turbulence in DUNE and HK (see Fig.~\ref{fig:var_cstar}).

The results presented in this paper rely on an effective two-flavour framework. This can be further improved to include three-flavour effects. The three-flavour scenario was considered in~\cite{Kneller:2014oea} where it was shown that the three flavours become important for turbulence amplitudes greater than $10$\% ($C_* >0.1$). Thus our approximation of using the effective two flavour scenario need to be re-examined for large turbulence amplitudes, which is beyond the scope of the current work.
Furthermore, we also did not consider the effect of collective flavour oscillations on our results. Particularly, if the neutrinos are susceptible to fast flavour instabilities occurring near the core, then flavour depolarization might already have happened by the time the neutrinos reach the shock-front~\cite{Bhattacharyya:2020dhu,Bhattacharyya:2022eed}. In that case, the turbulence signature becomes degenerate and cannot be disentangled from the depolarization due to fast conversions. 
Implementation of better statistical techniques including the comparison of binned and unbinned chi-squared analysis with improved treatment of reconstructed energy in the experiments and statistical priors, is beyond the scope of this work and is left for the future.

Understanding the complex nature of turbulence along with its very existence still remains a challenging problem. This work highlights the possibilities of constraining turbulence in the era of large-scale neutrino detectors like DUNE and HK. The statistics from a future galactic supernova at the upcoming neutrino detectors should be good enough to gain invaluable insights into turbulence. In fact, for a very nearby supernova (within $1$ kpc)~\cite{Mukhopadhyay:2020ubs,Healy:2023ovi} the complete spectrum can help place strong constraints on both the amplitude and the spectral index associated with turbulence, and allow us to investigate the double-dip in the neutrino spectra associated with shocks.  A planned $100$-kt neutrino detector THEIA~\cite{Theia:2019non} can also help place strong constraints owing to its large fiducial volume.  We conclude that with the next galactic core-collapse supernova in the era of DUNE and HK, we can significantly improve our understanding of the evolution of shocks inside the SN, and the underlying hydrodynamic instabilities. This can help to answer the question about whether the turbulence inside a core-collapse supernova follows a Kolmogorov spectrum.
\acknowledgments
We would like to thank Jose Carpio, Basudeb Dasgupta, Amol Dighe, Shigeo S. Kimura, Cecilia Lunardini, Kohta Murase, David Radice, and George Zahariade for helpful discussions and invaluable suggestions. M.\,M. wishes to thank the Astronomical Institute at Tohoku University, the Yukawa Institute for Theoretical Physics (YITP), Kyoto University, and the Institute for Advanced Study (IAS), Princeton, for their hospitality where major parts of this work were completed.
M.\,M. is supported by NSF Grant No. AST-2108466. M.\,M. also acknowledges support from the 
Institute for Gravitation and the Cosmos (IGC) Postdoctoral Fellowship. 
\appendix
\section{Algorithm}
\label{appsec:numerical_tech}
In this appendix, we outline the algorithm used for calculating the neutrino survival probabilities. Computing the spatial evolution of $\bm \Psi_e$ from Eq.~\eqref{eq:maineq} is enough to give us the survival and disappearance probabilities while the neutrinos propagate through some arbitrary matter profile. To do this we implement the following:
\begin{itemize}
\item Initial conditions: $L$, $n$ which specify $\Delta x$, $\theta$, $\Delta m^2$, $E$, and the initial wavefunction in the flavour basis $\bm \Psi_e (x_0)$. 
\item The information about the given matter density profile is taken into account through Eq.~\eqref{eq:effprof}.
\item We can then calculate $\theta_{M,i}$ and $\Delta m_{M,i}$ at the $i-$th slab using Eqs.~\eqref{eq:thetam} and~\eqref{eq:deltamsq} respectively.
\item Now, using Eqs.~\eqref{eq:unitev} and~\eqref{eq:unitar} we can evaluate the term in the square brackets which gives us the wavefunction $\bm \Psi (x_i)$ at the $i-$th slab, where, $i$ spans from $1,\dots,n$.
\item The survival or disappearance probability at every slab $i$ can then be calculated from Eq.~\eqref{eq:prob}. The mass eigenstate probabilities can be obtained by first getting the wavefunction in the mass basis using Eq.~\eqref{eq:masswvfn} and then calculating the probability.
\end{itemize}
\bibliography{refs}

\providecommand{\href}[2]{#2}\begingroup\raggedright\begin{thebibliography}{100}

\bibitem{Kamiokande-II:1987idp}
{\scshape Kamiokande-II} collaboration, \emph{{Observation of a Neutrino Burst from the Supernova SN 1987a}}, \href{https://doi.org/10.1103/PhysRevLett.58.1490}{\emph{Phys. Rev. Lett.} {\bfseries 58} (1987) 1490}.

\bibitem{Hirata:1988ad}
K.S.~Hirata et~al., \emph{{Observation in the Kamiokande-II Detector of the Neutrino Burst from Supernova SN 1987a}}, \href{https://doi.org/10.1103/PhysRevD.38.448}{\emph{Phys. Rev. D} {\bfseries 38} (1988) 448}.

\bibitem{Bionta:1987qt}
R.M.~Bionta et~al., \emph{{Observation of a Neutrino Burst in Coincidence with Supernova SN 1987a in the Large Magellanic Cloud}}, \href{https://doi.org/10.1103/PhysRevLett.58.1494}{\emph{Phys. Rev. Lett.} {\bfseries 58} (1987) 1494}.

\bibitem{Alekseev:1987ej}
E.N.~Alekseev, L.N.~Alekseeva, V.I.~Volchenko and I.V.~Krivosheina, \emph{{Possible Detection of a Neutrino Signal on 23 February 1987 at the Baksan Underground Scintillation Telescope of the Institute of Nuclear Research}}, {\emph{JETP Lett.} {\bfseries 45} (1987) 589}.

\bibitem{Duan:2005cp}
H.~Duan, G.M.~Fuller and Y.-Z.~Qian, \emph{{Collective neutrino flavor transformation in supernovae}}, \href{https://doi.org/10.1103/PhysRevD.74.123004}{\emph{Phys. Rev. D} {\bfseries 74} (2006) 123004} [\href{https://arxiv.org/abs/astro-ph/0511275}{{\ttfamily astro-ph/0511275}}].

\bibitem{Hannestad:2006nj}
S.~Hannestad, G.G.~Raffelt, G.~Sigl and Y.Y.Y.~Wong, \emph{{Self-induced conversion in dense neutrino gases: Pendulum in flavour space}}, \href{https://doi.org/10.1103/PhysRevD.74.105010, 10.1103/PhysRevD.76.029901}{\emph{Phys. Rev.} {\bfseries D74} (2006) 105010} [\href{https://arxiv.org/abs/astro-ph/0608695}{{\ttfamily astro-ph/0608695}}].

\bibitem{Duan:2007mv}
H.~Duan, G.M.~Fuller, J.~Carlson and Y.-Z.~Qian, \emph{{Analysis of Collective Neutrino Flavor Transformation in Supernovae}}, \href{https://doi.org/10.1103/PhysRevD.75.125005}{\emph{Phys. Rev. D} {\bfseries 75} (2007) 125005} [\href{https://arxiv.org/abs/astro-ph/0703776}{{\ttfamily astro-ph/0703776}}].

\bibitem{PhysRevD.83.105022}
G.G.~Raffelt, \emph{$n$-mode coherence in collective neutrino oscillations}, \href{https://doi.org/10.1103/PhysRevD.83.105022}{\emph{Phys. Rev. D} {\bfseries 83} (2011) 105022}.

\bibitem{Fogli_2007}
G.~Fogli, E.~Lisi, A.~Marrone and A.~Mirizzi, \emph{Collective neutrino flavor transitions in supernovae and the role of trajectory averaging}, \href{https://doi.org/10.1088/1475-7516/2007/12/010}{\emph{Journal of Cosmology and Astroparticle Physics} {\bfseries 2007} (2007) 010}.

\bibitem{Raffelt:2007yz}
G.G.~Raffelt and G.~Sigl, \emph{{Self-induced decoherence in dense neutrino gases}}, \href{https://doi.org/10.1103/PhysRevD.75.083002}{\emph{Phys. Rev. D} {\bfseries 75} (2007) 083002} [\href{https://arxiv.org/abs/hep-ph/0701182}{{\ttfamily hep-ph/0701182}}].

\bibitem{Esteban-Pretel:2007jwl}
A.~Esteban-Pretel, S.~Pastor, R.~Tomas, G.G.~Raffelt and G.~Sigl, \emph{{Decoherence in supernova neutrino transformations suppressed by deleptonization}}, \href{https://doi.org/10.1103/PhysRevD.76.125018}{\emph{Phys. Rev. D} {\bfseries 76} (2007) 125018} [\href{https://arxiv.org/abs/0706.2498}{{\ttfamily 0706.2498}}].

\bibitem{Dasgupta:2007ws}
B.~Dasgupta and A.~Dighe, \emph{{Collective three-flavor oscillations of supernova neutrinos}}, \href{https://doi.org/10.1103/PhysRevD.77.113002}{\emph{Phys. Rev. D} {\bfseries 77} (2008) 113002} [\href{https://arxiv.org/abs/0712.3798}{{\ttfamily 0712.3798}}].

\bibitem{Dasgupta:2008my}
B.~Dasgupta, A.~Dighe and A.~Mirizzi, \emph{{Identifying neutrino mass hierarchy at extremely small theta(13) through Earth matter effects in a supernova signal}}, \href{https://doi.org/10.1103/PhysRevLett.101.171801}{\emph{Phys. Rev. Lett.} {\bfseries 101} (2008) 171801} [\href{https://arxiv.org/abs/0802.1481}{{\ttfamily 0802.1481}}].

\bibitem{Dasgupta:2010cd}
B.~Dasgupta, A.~Mirizzi, I.~Tamborra and R.~Tomas, \emph{{Neutrino mass hierarchy and three-flavor spectral splits of supernova neutrinos}}, \href{https://doi.org/10.1103/PhysRevD.81.093008}{\emph{Phys. Rev. D} {\bfseries 81} (2010) 093008} [\href{https://arxiv.org/abs/1002.2943}{{\ttfamily 1002.2943}}].

\bibitem{Raffelt:2007cb}
G.G.~Raffelt and A.Y.~Smirnov, \emph{{Self-induced spectral splits in supernova neutrino fluxes}}, \href{https://doi.org/10.1103/PhysRevD.76.081301, 10.1103/PhysRevD.77.029903}{\emph{Phys. Rev.} {\bfseries D76} (2007) 081301} [\href{https://arxiv.org/abs/0705.1830}{{\ttfamily 0705.1830}}].

\bibitem{Sarikas:2011am}
S.~Sarikas, G.G.~Raffelt, L.~Hudepohl and H.-T.~Janka, \emph{{Suppression of Self-Induced Flavor Conversion in the Supernova Accretion Phase}}, \href{https://doi.org/10.1103/PhysRevLett.108.061101}{\emph{Phys. Rev. Lett.} {\bfseries 108} (2012) 061101} [\href{https://arxiv.org/abs/1109.3601}{{\ttfamily 1109.3601}}].

\bibitem{Banerjee:2011fj}
A.~Banerjee, A.~Dighe and G.G.~Raffelt, \emph{{Linearized flavor-stability analysis of dense neutrino streams}}, \href{https://doi.org/10.1103/PhysRevD.84.053013}{\emph{Phys. Rev.} {\bfseries D84} (2011) 053013} [\href{https://arxiv.org/abs/1107.2308}{{\ttfamily 1107.2308}}].

\bibitem{Raffelt:2013rqa}
G.~Raffelt, S.~Sarikas and D.~de~Sousa~Seixas, \emph{{Axial Symmetry Breaking in Self-Induced Flavor Conversion of Supernova Neutrino Fluxes}}, \href{https://doi.org/10.1103/PhysRevLett.111.091101}{\emph{Phys. Rev. Lett.} {\bfseries 111} (2013) 091101} [\href{https://arxiv.org/abs/1305.7140}{{\ttfamily 1305.7140}}].

\bibitem{Chakraborty:2015tfa}
S.~Chakraborty, R.S.~Hansen, I.~Izaguirre and G.~Raffelt, \emph{{Self-induced flavor conversion of supernova neutrinos on small scales}}, \href{https://doi.org/10.1088/1475-7516/2016/01/028}{\emph{JCAP} {\bfseries 01} (2016) 028} [\href{https://arxiv.org/abs/1507.07569}{{\ttfamily 1507.07569}}].

\bibitem{Mirizzi:2012wp}
A.~Mirizzi and P.D.~Serpico, \emph{{Flavor Stability Analysis of Dense Supernova Neutrinos with Flavor-Dependent Angular Distributions}}, \href{https://doi.org/10.1103/PhysRevD.86.085010}{\emph{Phys. Rev. D} {\bfseries 86} (2012) 085010} [\href{https://arxiv.org/abs/1208.0157}{{\ttfamily 1208.0157}}].

\bibitem{Capozzi:2016oyk}
F.~Capozzi, B.~Dasgupta and A.~Mirizzi, \emph{{Self-induced temporal instability from a neutrino antenna}}, \href{https://doi.org/10.1088/1475-7516/2016/04/043}{\emph{JCAP} {\bfseries 1604} (2016) 043} [\href{https://arxiv.org/abs/1603.03288}{{\ttfamily 1603.03288}}].

\bibitem{Das:2017iuj}
A.~Das, A.~Dighe and M.~Sen, \emph{{New effects of non-standard self-interactions of neutrinos in a supernova}}, \href{https://doi.org/10.1088/1475-7516/2017/05/051}{\emph{JCAP} {\bfseries 05} (2017) 051} [\href{https://arxiv.org/abs/1705.00468}{{\ttfamily 1705.00468}}].

\bibitem{Airen:2018nvp}
S.~Airen, F.~Capozzi, S.~Chakraborty, B.~Dasgupta, G.~Raffelt and T.~Stirner, \emph{{Normal-mode Analysis for Collective Neutrino Oscillations}}, \href{https://doi.org/10.1088/1475-7516/2018/12/019}{\emph{JCAP} {\bfseries 12} (2018) 019} [\href{https://arxiv.org/abs/1809.09137}{{\ttfamily 1809.09137}}].

\bibitem{Stapleford:2019yqg}
C.J.~Stapleford, C.~Fr\"ohlich and J.P.~Kneller, \emph{{Coupling Neutrino Oscillations and Simulations of Core-Collapse Supernovae}}, \href{https://doi.org/10.1103/PhysRevD.102.081301}{\emph{Phys. Rev. D} {\bfseries 102} (2020) 081301} [\href{https://arxiv.org/abs/1910.04172}{{\ttfamily 1910.04172}}].

\bibitem{Sawyer:2015dsa}
R.F.~Sawyer, \emph{{Neutrino cloud instabilities just above the neutrino sphere of a supernova}}, \href{https://doi.org/10.1103/PhysRevLett.116.081101}{\emph{Phys. Rev. Lett.} {\bfseries 116} (2016) 081101} [\href{https://arxiv.org/abs/1509.03323}{{\ttfamily 1509.03323}}].

\bibitem{Chakraborty:2016lct}
S.~Chakraborty, R.S.~Hansen, I.~Izaguirre and G.G.~Raffelt, \emph{{Self-induced neutrino flavor conversion without flavor mixing}}, \href{https://doi.org/10.1088/1475-7516/2016/03/042}{\emph{JCAP} {\bfseries 1603} (2016) 042} [\href{https://arxiv.org/abs/1602.00698}{{\ttfamily 1602.00698}}].

\bibitem{Dasgupta:2016dbv}
B.~Dasgupta, A.~Mirizzi and M.~Sen, \emph{{Fast neutrino flavor conversions near the supernova core with realistic flavor-dependent angular distributions}}, \href{https://doi.org/10.1088/1475-7516/2017/02/019}{\emph{JCAP} {\bfseries 1702} (2017) 019} [\href{https://arxiv.org/abs/1609.00528}{{\ttfamily 1609.00528}}].

\bibitem{Sawyer:2005jk}
R.F.~Sawyer, \emph{{Speed-up of neutrino transformations in a supernova environment}}, \href{https://doi.org/10.1103/PhysRevD.72.045003}{\emph{Phys. Rev.} {\bfseries D72} (2005) 045003} [\href{https://arxiv.org/abs/hep-ph/0503013}{{\ttfamily hep-ph/0503013}}].

\bibitem{Sawyer:2008zs}
R.F.~Sawyer, \emph{{The multi-angle instability in dense neutrino systems}}, \href{https://doi.org/10.1103/PhysRevD.79.105003}{\emph{Phys. Rev.} {\bfseries D79} (2009) 105003} [\href{https://arxiv.org/abs/0803.4319}{{\ttfamily 0803.4319}}].

\bibitem{Izaguirre:2016gsx}
I.~Izaguirre, G.G.~Raffelt and I.~Tamborra, \emph{{Fast Pairwise Conversion of Supernova Neutrinos: A Dispersion-Relation Approach}}, \href{https://doi.org/10.1103/PhysRevLett.118.021101}{\emph{Phys. Rev. Lett.} {\bfseries 118} (2017) 021101} [\href{https://arxiv.org/abs/1610.01612}{{\ttfamily 1610.01612}}].

\bibitem{Capozzi:2017gqd}
F.~Capozzi, B.~Dasgupta, E.~Lisi, A.~Marrone and A.~Mirizzi, \emph{{Fast flavor conversions of supernova neutrinos: Classifying instabilities via dispersion relations}}, \href{https://doi.org/10.1103/PhysRevD.96.043016}{\emph{Phys. Rev.} {\bfseries D96} (2017) 043016} [\href{https://arxiv.org/abs/1706.03360}{{\ttfamily 1706.03360}}].

\bibitem{Dighe:2017sur}
A.~Dighe and M.~Sen, \emph{{Nonstandard neutrino self-interactions in a supernova and fast flavor conversions}}, \href{https://doi.org/10.1103/PhysRevD.97.043011}{\emph{Phys. Rev. D} {\bfseries 97} (2018) 043011} [\href{https://arxiv.org/abs/1709.06858}{{\ttfamily 1709.06858}}].

\bibitem{Dasgupta:2017oko}
B.~Dasgupta and M.~Sen, \emph{{Fast Neutrino Flavor Conversion as Oscillations in a Quartic Potential}}, \href{https://doi.org/10.1103/PhysRevD.97.023017}{\emph{Phys. Rev.} {\bfseries D97} (2018) 023017} [\href{https://arxiv.org/abs/1709.08671}{{\ttfamily 1709.08671}}].

\bibitem{Dasgupta:2018ulw}
B.~Dasgupta, A.~Mirizzi and M.~Sen, \emph{{Simple method of diagnosing fast flavor conversions of supernova neutrinos}}, \href{https://doi.org/10.1103/PhysRevD.98.103001}{\emph{Phys. Rev.} {\bfseries D98} (2018) 103001} [\href{https://arxiv.org/abs/1807.03322}{{\ttfamily 1807.03322}}].

\bibitem{Abbar:2018shq}
S.~Abbar, H.~Duan, K.~Sumiyoshi, T.~Takiwaki and M.C.~Volpe, \emph{{On the occurrence of fast neutrino flavor conversions in multidimensional supernova models}},  \href{https://arxiv.org/abs/1812.06883}{{\ttfamily 1812.06883}}.

\bibitem{Azari:2019jvr}
M.D.~Azari, S.~Yamada, T.~Morinaga, W.~Iwakami, H.~Nagakura and K.~Sumiyoshi, \emph{{Linear Analysis of Fast-Pairwise Collective Neutrino Oscillations in Core-Collapse Supernovae based on the Results of Boltzmann Simulations}},  \href{https://arxiv.org/abs/1902.07467}{{\ttfamily 1902.07467}}.

\bibitem{Johns:2019izj}
L.~Johns, H.~Nagakura, G.M.~Fuller and A.~Burrows, \emph{{Neutrino oscillations in supernovae: angular moments and fast instabilities}}, \href{https://doi.org/10.1103/PhysRevD.101.043009}{\emph{Phys. Rev. D} {\bfseries 101} (2020) 043009} [\href{https://arxiv.org/abs/1910.05682}{{\ttfamily 1910.05682}}].

\bibitem{Glas:2019ijo}
R.~Glas, H.T.~Janka, F.~Capozzi, M.~Sen, B.~Dasgupta, A.~Mirizzi et~al., \emph{{Fast Neutrino Flavor Instability in the Neutron-star Convection Layer of Three-dimensional Supernova Models}}, \href{https://doi.org/10.1103/PhysRevD.101.063001}{\emph{Phys. Rev. D} {\bfseries 101} (2020) 063001} [\href{https://arxiv.org/abs/1912.00274}{{\ttfamily 1912.00274}}].

\bibitem{Shalgar:2019qwg}
S.~Shalgar, I.~Padilla-Gay and I.~Tamborra, \emph{{Neutrino propagation hinders fast pairwise flavor conversions}},  \href{https://arxiv.org/abs/1911.09110}{{\ttfamily 1911.09110}}.

\bibitem{Abbar:2020fcl}
S.~Abbar, \emph{{Searching for Fast Neutrino Flavor Conversion Modes in Core-collapse Supernova Simulations}}, \href{https://doi.org/10.1088/1475-7516/2020/05/027}{\emph{JCAP} {\bfseries 05} (2020) 027} [\href{https://arxiv.org/abs/2003.00969}{{\ttfamily 2003.00969}}].

\bibitem{Bhattacharyya:2020dhu}
S.~Bhattacharyya and B.~Dasgupta, \emph{{Fast Neutrino Flavor Conversion at Late Time}},  \href{https://arxiv.org/abs/2005.00459}{{\ttfamily 2005.00459}}.

\bibitem{Bhattacharyya:2020jpj}
S.~Bhattacharyya and B.~Dasgupta, \emph{{Fast Flavor Depolarization of Supernova Neutrinos}}, \href{https://doi.org/10.1103/PhysRevLett.126.061302}{\emph{Phys. Rev. Lett.} {\bfseries 126} (2021) 061302} [\href{https://arxiv.org/abs/2009.03337}{{\ttfamily 2009.03337}}].

\bibitem{Li:2021vqj}
X.~Li and D.M.~Siegel, \emph{{Neutrino Fast Flavor Conversions in Neutron-Star Postmerger Accretion Disks}}, \href{https://doi.org/10.1103/PhysRevLett.126.251101}{\emph{Phys. Rev. Lett.} {\bfseries 126} (2021) 251101} [\href{https://arxiv.org/abs/2103.02616}{{\ttfamily 2103.02616}}].

\bibitem{Padilla-Gay:2021haz}
I.~Padilla-Gay, I.~Tamborra and G.G.~Raffelt, \emph{{Neutrino Flavor Pendulum Reloaded: The Case of Fast Pairwise Conversion}}, \href{https://doi.org/10.1103/PhysRevLett.128.121102}{\emph{Phys. Rev. Lett.} {\bfseries 128} (2022) 121102} [\href{https://arxiv.org/abs/2109.14627}{{\ttfamily 2109.14627}}].

\bibitem{Wu:2021uvt}
M.-R.~Wu, M.~George, C.-Y.~Lin and Z.~Xiong, \emph{{Collective fast neutrino flavor conversions in a 1D box: Initial conditions and long-term evolution}}, \href{https://doi.org/10.1103/PhysRevD.104.103003}{\emph{Phys. Rev. D} {\bfseries 104} (2021) 103003} [\href{https://arxiv.org/abs/2108.09886}{{\ttfamily 2108.09886}}].

\bibitem{Richers:2021nbx}
S.~Richers, D.E.~Willcox, N.M.~Ford and A.~Myers, \emph{{Particle-in-cell Simulation of the Neutrino Fast Flavor Instability}}, \href{https://doi.org/10.1103/PhysRevD.103.083013}{\emph{Phys. Rev. D} {\bfseries 103} (2021) 083013} [\href{https://arxiv.org/abs/2101.02745}{{\ttfamily 2101.02745}}].

\bibitem{Richers:2021xtf}
S.~Richers, D.~Willcox and N.~Ford, \emph{{Neutrino fast flavor instability in three dimensions}}, \href{https://doi.org/10.1103/PhysRevD.104.103023}{\emph{Phys. Rev. D} {\bfseries 104} (2021) 103023} [\href{https://arxiv.org/abs/2109.08631}{{\ttfamily 2109.08631}}].

\bibitem{Martin:2019gxb}
J.D.~Martin, C.~Yi and H.~Duan, \emph{{Dynamic fast flavor oscillation waves in dense neutrino gases}}, \href{https://doi.org/10.1016/j.physletb.2019.135088}{\emph{Phys. Lett. B} {\bfseries 800} (2020) 135088} [\href{https://arxiv.org/abs/1909.05225}{{\ttfamily 1909.05225}}].

\bibitem{Abbar:2018beu}
S.~Abbar and M.C.~Volpe, \emph{{On Fast Neutrino Flavor Conversion Modes in the Nonlinear Regime}}, \href{https://doi.org/10.1016/j.physletb.2019.02.002}{\emph{Phys. Lett.} {\bfseries B790} (2019) 545} [\href{https://arxiv.org/abs/1811.04215}{{\ttfamily 1811.04215}}].

\bibitem{Johns:2020qsk}
L.~Johns, H.~Nagakura, G.M.~Fuller and A.~Burrows, \emph{{Fast oscillations, collisionless relaxation, and spurious evolution of supernova neutrino flavor}}, \href{https://doi.org/10.1103/PhysRevD.102.103017}{\emph{Phys. Rev. D} {\bfseries 102} (2020) 103017} [\href{https://arxiv.org/abs/2009.09024}{{\ttfamily 2009.09024}}].

\bibitem{Sigl:2021tmj}
G.~Sigl, \emph{{Simulations of fast neutrino flavor conversions with interactions in inhomogeneous media}}, \href{https://doi.org/10.1103/PhysRevD.105.043005}{\emph{Phys. Rev. D} {\bfseries 105} (2022) 043005} [\href{https://arxiv.org/abs/2109.00091}{{\ttfamily 2109.00091}}].

\bibitem{Xiong:2021dex}
Z.~Xiong and Y.-Z.~Qian, \emph{{Stationary solutions for fast flavor oscillations of a homogeneous dense neutrino gas}}, \href{https://doi.org/10.1016/j.physletb.2021.136550}{\emph{Phys. Lett. B} {\bfseries 820} (2021) 136550} [\href{https://arxiv.org/abs/2104.05618}{{\ttfamily 2104.05618}}].

\bibitem{Abbar:2021lmm}
S.~Abbar and F.~Capozzi, \emph{{Suppression of fast neutrino flavor conversions occurring at large distances in core-collapse supernovae}}, \href{https://doi.org/10.1088/1475-7516/2022/03/051}{\emph{JCAP} {\bfseries 03} (2022) 051} [\href{https://arxiv.org/abs/2111.14880}{{\ttfamily 2111.14880}}].

\bibitem{DelfanAzari:2019tez}
M.~Delfan~Azari, S.~Yamada, T.~Morinaga, H.~Nagakura, S.~Furusawa, A.~Harada et~al., \emph{{Fast collective neutrino oscillations inside the neutrino sphere in core-collapse supernovae}}, \href{https://doi.org/10.1103/PhysRevD.101.023018}{\emph{Phys. Rev. D} {\bfseries 101} (2020) 023018} [\href{https://arxiv.org/abs/1910.06176}{{\ttfamily 1910.06176}}].

\bibitem{Chakraborty:2019wxe}
M.~Chakraborty and S.~Chakraborty, \emph{{Three flavor neutrino conversions in supernovae: slow \& fast instabilities}}, \href{https://doi.org/10.1088/1475-7516/2020/01/005}{\emph{JCAP} {\bfseries 2001} (2020) 005} [\href{https://arxiv.org/abs/1909.10420}{{\ttfamily 1909.10420}}].

\bibitem{Capozzi:2020kge}
F.~Capozzi, M.~Chakraborty, S.~Chakraborty and M.~Sen, \emph{{Fast flavor conversions in supernovae: the rise of mu-tau neutrinos}}, \href{https://doi.org/10.1103/PhysRevLett.125.251801}{\emph{Phys. Rev. Lett.} {\bfseries 125} (2020) 251801} [\href{https://arxiv.org/abs/2005.14204}{{\ttfamily 2005.14204}}].

\bibitem{Bhattacharyya:2021klg}
S.~Bhattacharyya and B.~Dasgupta, \emph{{Fast flavor oscillations of astrophysical neutrinos with 1, 2, \textellipsis{}, \ensuremath{\infty} crossings}}, \href{https://doi.org/10.1088/1475-7516/2021/07/023}{\emph{JCAP} {\bfseries 07} (2021) 023} [\href{https://arxiv.org/abs/2101.01226}{{\ttfamily 2101.01226}}].

\bibitem{Nagakura:2021suv}
H.~Nagakura and L.~Johns, \emph{{New method for detecting fast neutrino flavor conversions in core-collapse supernova models with two-moment neutrino transport}}, \href{https://doi.org/10.1103/PhysRevD.104.063014}{\emph{Phys. Rev. D} {\bfseries 104} (2021) 063014} [\href{https://arxiv.org/abs/2106.02650}{{\ttfamily 2106.02650}}].

\bibitem{Nagakura:2021hyb}
H.~Nagakura, L.~Johns, A.~Burrows and G.M.~Fuller, \emph{{Where, when, and why: Occurrence of fast-pairwise collective neutrino oscillation in three-dimensional core-collapse supernova models}}, \href{https://doi.org/10.1103/PhysRevD.104.083025}{\emph{Phys. Rev. D} {\bfseries 104} (2021) 083025} [\href{https://arxiv.org/abs/2108.07281}{{\ttfamily 2108.07281}}].

\bibitem{Dasgupta:2021gfs}
B.~Dasgupta, \emph{{Collective Neutrino Flavor Instability Requires a Crossing}}, \href{https://doi.org/10.1103/PhysRevLett.128.081102}{\emph{Phys. Rev. Lett.} {\bfseries 128} (2022) 081102} [\href{https://arxiv.org/abs/2110.00192}{{\ttfamily 2110.00192}}].

\bibitem{Shalgar:2021wlj}
S.~Shalgar and I.~Tamborra, \emph{{Three flavor revolution in fast pairwise neutrino conversion}}, \href{https://doi.org/10.1103/PhysRevD.104.023011}{\emph{Phys. Rev. D} {\bfseries 104} (2021) 023011} [\href{https://arxiv.org/abs/2103.12743}{{\ttfamily 2103.12743}}].

\bibitem{Bhattacharyya:2022eed}
S.~Bhattacharyya and B.~Dasgupta, \emph{{Elaborating the Ultimate Fate of Fast Collective Neutrino Flavor Oscillations}},  \href{https://arxiv.org/abs/2205.05129}{{\ttfamily 2205.05129}}.

\bibitem{Wu:2017drk}
M.-R.~Wu, I.~Tamborra, O.~Just and H.-T.~Janka, \emph{{Imprints of neutrino-pair flavor conversions on nucleosynthesis in ejecta from neutron-star merger remnants}}, \href{https://doi.org/10.1103/PhysRevD.96.123015}{\emph{Phys. Rev. D} {\bfseries 96} (2017) 123015} [\href{https://arxiv.org/abs/1711.00477}{{\ttfamily 1711.00477}}].

\bibitem{Zaizen:2019ufj}
M.~Zaizen, J.F.~Cherry, T.~Takiwaki, S.~Horiuchi, K.~Kotake, H.~Umeda et~al., \emph{{Neutrino halo effect on collective neutrino oscillation in iron core-collapse supernova model of a 9.6 $M_{\odot}$ star}}, \href{https://doi.org/10.1088/1475-7516/2020/06/011}{\emph{JCAP} {\bfseries 06} (2020) 011} [\href{https://arxiv.org/abs/1908.10594}{{\ttfamily 1908.10594}}].

\bibitem{Nagakura:2019sig}
H.~Nagakura, T.~Morinaga, C.~Kato and S.~Yamada, \emph{{Fast-pairwise collective neutrino oscillations associated with asymmetric neutrino emissions in core-collapse supernova}},  \href{https://arxiv.org/abs/1910.04288}{{\ttfamily 1910.04288}}.

\bibitem{Duan:2010bg}
H.~Duan, G.M.~Fuller and Y.-Z.~Qian, \emph{Collective neutrino oscillations}, \href{https://doi.org/10.1146/annurev.nucl.012809.104524}{\emph{Ann. Rev. Nucl. Part. Sci.} {\bfseries 60} (2010) 569} [\href{https://arxiv.org/abs/1001.2799}{{\ttfamily 1001.2799}}].

\bibitem{Mirizzi:2015eza}
A.~Mirizzi, I.~Tamborra, H.-T.~Janka, N.~Saviano, K.~Scholberg, R.~Bollig et~al., \emph{{Supernova Neutrinos: Production, Oscillations and Detection}}, \href{https://doi.org/10.1393/ncr/i2016-10120-8}{\emph{Riv. Nuovo Cim.} {\bfseries 39} (2016) 1} [\href{https://arxiv.org/abs/1508.00785}{{\ttfamily 1508.00785}}].

\bibitem{Tamborra:2020cul}
I.~Tamborra and S.~Shalgar, \emph{{New Developments in Flavor Evolution of a Dense Neutrino Gas}}, \href{https://doi.org/10.1146/annurev-nucl-102920-050505}{\emph{Ann. Rev. Nucl. Part. Sci.} {\bfseries 71} (2021) 165} [\href{https://arxiv.org/abs/2011.01948}{{\ttfamily 2011.01948}}].

\bibitem{Richers:2022zug}
S.~Richers and M.~Sen, \emph{{Fast Flavor Transformations}},  in \emph{{Handbook of Nuclear Physics}}, I.~Tanihata, H.~Toki and T.~Kajino, eds., pp.~1--17 (2022), \href{https://doi.org/10.1007/978-981-15-8818-1_125-1}{DOI} [\href{https://arxiv.org/abs/2207.03561}{{\ttfamily 2207.03561}}].

\bibitem{Mikheev:1986gs}
S.P.~Mikheev and A.Y.~Smirnov, \emph{{Resonance Amplification of Oscillations in Matter and Spectroscopy of Solar Neutrinos}}, {\emph{Sov. J. Nucl. Phys.} {\bfseries 42} (1985) 913}.

\bibitem{PhysRevD.17.2369}
L.~Wolfenstein, \emph{Neutrino oscillations in matter}, \href{https://doi.org/10.1103/PhysRevD.17.2369}{\emph{Phys. Rev. D} {\bfseries 17} (1978) 2369}.

\bibitem{Schirato:2002tg}
R.C.~Schirato and G.M.~Fuller, \emph{{Connection between supernova shocks, flavor transformation, and the neutrino signal}},  \href{https://arxiv.org/abs/astro-ph/0205390}{{\ttfamily astro-ph/0205390}}.

\bibitem{Takahashi:2002yj}
K.~Takahashi, K.~Sato, H.E.~Dalhed and J.R.~Wilson, \emph{{Shock propagation and neutrino oscillation in supernova}}, \href{https://doi.org/10.1016/S0927-6505(03)00175-0}{\emph{Astropart. Phys.} {\bfseries 20} (2003) 189} [\href{https://arxiv.org/abs/astro-ph/0212195}{{\ttfamily astro-ph/0212195}}].

\bibitem{Lunardini:2003eh}
C.~Lunardini and A.Y.~Smirnov, \emph{{Probing the neutrino mass hierarchy and the 13 mixing with supernovae}}, \href{https://doi.org/10.1088/1475-7516/2003/06/009}{\emph{JCAP} {\bfseries 06} (2003) 009} [\href{https://arxiv.org/abs/hep-ph/0302033}{{\ttfamily hep-ph/0302033}}].

\bibitem{Fogli:2003dw}
G.L.~Fogli, E.~Lisi, D.~Montanino and A.~Mirizzi, \emph{{Analysis of energy and time dependence of supernova shock effects on neutrino crossing probabilities}}, \href{https://doi.org/10.1103/PhysRevD.68.033005}{\emph{Phys. Rev.} {\bfseries D68} (2003) 033005} [\href{https://arxiv.org/abs/hep-ph/0304056}{{\ttfamily hep-ph/0304056}}].

\bibitem{Fogli:2004ff}
G.L.~Fogli, E.~Lisi, A.~Mirizzi and D.~Montanino, \emph{{Probing supernova shock waves and neutrino flavor transitions in next-generation water-Cerenkov detectors}}, \href{https://doi.org/10.1088/1475-7516/2005/04/002}{\emph{JCAP} {\bfseries 0504} (2005) 002} [\href{https://arxiv.org/abs/hep-ph/0412046}{{\ttfamily hep-ph/0412046}}].

\bibitem{Tomas:2004gr}
R.~Tomas, M.~Kachelriess, G.~Raffelt, A.~Dighe, H.T.~Janka and L.~Scheck, \emph{{Neutrino signatures of supernova shock and reverse shock propagation}}, \href{https://doi.org/10.1088/1475-7516/2004/09/015}{\emph{JCAP} {\bfseries 09} (2004) 015} [\href{https://arxiv.org/abs/astro-ph/0407132}{{\ttfamily astro-ph/0407132}}].

\bibitem{Dasgupta:2005wn}
B.~Dasgupta and A.~Dighe, \emph{{Phase effects in neutrino conversions during a supernova shock wave}}, \href{https://doi.org/10.1103/PhysRevD.75.093002}{\emph{Phys. Rev. D} {\bfseries 75} (2007) 093002} [\href{https://arxiv.org/abs/hep-ph/0510219}{{\ttfamily hep-ph/0510219}}].

\bibitem{Galais:2009wi}
S.~Galais, J.~Kneller, C.~Volpe and J.~Gava, \emph{{Shockwaves in Supernovae: New Implications on the Diffuse Supernova Neutrino Background}}, \href{https://doi.org/10.1103/PhysRevD.81.053002}{\emph{Phys. Rev.} {\bfseries D81} (2010) 053002} [\href{https://arxiv.org/abs/0906.5294}{{\ttfamily 0906.5294}}].

\bibitem{Friedland:2020ecy}
A.~Friedland and P.~Mukhopadhyay, \emph{{Near-critical supernova outflows and their neutrino signatures}}, \href{https://doi.org/10.1016/j.physletb.2022.137403}{\emph{Phys. Lett. B} {\bfseries 834} (2022) 137403} [\href{https://arxiv.org/abs/2009.10059}{{\ttfamily 2009.10059}}].

\bibitem{Radice:2017kmj}
D.~Radice, E.~Abdikamalov, C.D.~Ott, P.~M\"osta, S.M.~Couch and L.F.~Roberts, \emph{{Turbulence in Core-Collapse Supernovae}}, \href{https://doi.org/10.1088/1361-6471/aab872}{\emph{J. Phys. G} {\bfseries 45} (2018) 053003} [\href{https://arxiv.org/abs/1710.01282}{{\ttfamily 1710.01282}}].

\bibitem{Fogli:2006xy}
G.L.~Fogli, E.~Lisi, A.~Mirizzi and D.~Montanino, \emph{{Damping of supernova neutrino transitions in stochastic shock-wave density profiles}}, \href{https://doi.org/10.1088/1475-7516/2006/06/012}{\emph{JCAP} {\bfseries 06} (2006) 012} [\href{https://arxiv.org/abs/hep-ph/0603033}{{\ttfamily hep-ph/0603033}}].

\bibitem{friedland2006neutrino}
A.~Friedland and A.~Gruzinov, \emph{Neutrino signatures of supernova turbulence}, {\emph{arXiv preprint astro-ph/0607244} (2006) } [\href{https://arxiv.org/abs/astro-ph/0607244}{{\ttfamily astro-ph/0607244}}].

\bibitem{Kneller:2010sc}
J.P.~Kneller and C.~Volpe, \emph{{Turbulence effects on supernova neutrinos}}, \href{https://doi.org/10.1103/PhysRevD.82.123004}{\emph{Phys. Rev. D} {\bfseries 82} (2010) 123004} [\href{https://arxiv.org/abs/1006.0913}{{\ttfamily 1006.0913}}].

\bibitem{Borriello:2013tha}
E.~Borriello, S.~Chakraborty, H.-T.~Janka, E.~Lisi and A.~Mirizzi, \emph{{Turbulence patterns and neutrino flavor transitions in high-resolution supernova models}}, \href{https://doi.org/10.1088/1475-7516/2014/11/030}{\emph{JCAP} {\bfseries 11} (2014) 030} [\href{https://arxiv.org/abs/1310.7488}{{\ttfamily 1310.7488}}].

\bibitem{Lund:2013uta}
T.~Lund and J.P.~Kneller, \emph{{Combining collective, MSW, and turbulence effects in supernova neutrino flavor evolution}}, \href{https://doi.org/10.1103/PhysRevD.88.023008}{\emph{Phys. Rev. D} {\bfseries 88} (2013) 023008} [\href{https://arxiv.org/abs/1304.6372}{{\ttfamily 1304.6372}}].

\bibitem{Kneller:2014oea}
J.P.~Kneller and N.V.~Kabadi, \emph{{Sensitivity of neutrinos to the supernova turbulence power spectrum: Point source statistics}}, \href{https://doi.org/10.1103/PhysRevD.92.013009}{\emph{Phys. Rev. D} {\bfseries 92} (2015) 013009} [\href{https://arxiv.org/abs/1410.5698}{{\ttfamily 1410.5698}}].

\bibitem{Patton:2014lza}
K.M.~Patton, J.P.~Kneller and G.C.~McLaughlin, \emph{{Stimulated neutrino transformation through turbulence on a changing density profile and application to supernovae}}, \href{https://doi.org/10.1103/PhysRevD.91.025001}{\emph{Phys. Rev. D} {\bfseries 91} (2015) 025001} [\href{https://arxiv.org/abs/1407.7835}{{\ttfamily 1407.7835}}].

\bibitem{Yang:2015oya}
Y.~Yang and J.P.~Kneller, \emph{{Neutrino Flavour Evolution Through Fluctuating Matter}}, \href{https://doi.org/10.1088/1361-6471/aab0c4}{\emph{J. Phys. G} {\bfseries 45} (2018) 045201} [\href{https://arxiv.org/abs/1510.01998}{{\ttfamily 1510.01998}}].

\bibitem{Kneller:2017lqg}
J.P.~Kneller and M.~de~los Reyes, \emph{{The effect of core-collapse supernova accretion phase turbulence on neutrino flavor evolution}}, \href{https://doi.org/10.1088/1361-6471/aa7bc8}{\emph{J. Phys. G} {\bfseries 44} (2017) 084008} [\href{https://arxiv.org/abs/1702.06951}{{\ttfamily 1702.06951}}].

\bibitem{Abbar:2020ror}
S.~Abbar, \emph{{Turbulence Fingerprint on Collective Oscillations of Supernova Neutrinos}}, \href{https://doi.org/10.1103/PhysRevD.103.045014}{\emph{Phys. Rev. D} {\bfseries 103} (2021) 045014} [\href{https://arxiv.org/abs/2007.13655}{{\ttfamily 2007.13655}}].

\bibitem{Dolence:2012kh}
J.C.~Dolence, A.~Burrows, J.W.~Murphy and J.~Nordhaus, \emph{{Dimensional Dependence of the Hydrodynamics of Core-Collapse Supernovae}}, \href{https://doi.org/10.1088/0004-637X/765/2/110}{\emph{Astrophys. J.} {\bfseries 765} (2013) 110} [\href{https://arxiv.org/abs/1210.5241}{{\ttfamily 1210.5241}}].

\bibitem{Couch:2014kza}
S.M.~Couch and C.D.~Ott, \emph{{The Role of Turbulence in Neutrino-Driven Core-Collapse Supernova Explosions}}, \href{https://doi.org/10.1088/0004-637X/799/1/5}{\emph{Astrophys. J.} {\bfseries 799} (2015) 5} [\href{https://arxiv.org/abs/1408.1399}{{\ttfamily 1408.1399}}].

\bibitem{Kneller:2007kg}
J.P.~Kneller, G.C.~McLaughlin and J.~Brockman, \emph{{Oscillation Effects and Time Variation of the Supernova Neutrino Signal}}, \href{https://doi.org/10.1103/PhysRevD.77.045023}{\emph{Phys. Rev. D} {\bfseries 77} (2008) 045023} [\href{https://arxiv.org/abs/0705.3835}{{\ttfamily 0705.3835}}].

\bibitem{PhysRevD.35.896}
F.J.~Botella, C.S.~Lim and W.J.~Marciano, \emph{Radiative corrections to neutrino indices of refraction}, \href{https://doi.org/10.1103/PhysRevD.35.896}{\emph{Phys. Rev. D} {\bfseries 35} (1987) 896}.

\bibitem{Akhmedov:2002zj}
E.K.~Akhmedov, C.~Lunardini and A.Y.~Smirnov, \emph{{Supernova neutrinos: Difference of muon-neutrino - tau-neutrino fluxes and conversion effects}}, \href{https://doi.org/10.1016/S0550-3213(02)00692-2}{\emph{Nucl. Phys. B} {\bfseries 643} (2002) 339} [\href{https://arxiv.org/abs/hep-ph/0204091}{{\ttfamily hep-ph/0204091}}].

\bibitem{Mirizzi:2009td}
A.~Mirizzi, S.~Pozzorini, G.G.~Raffelt and P.D.~Serpico, \emph{{Flavour-dependent radiative correction to neutrino-neutrino refraction}}, \href{https://doi.org/10.1088/1126-6708/2009/10/020}{\emph{JHEP} {\bfseries 10} (2009) 020} [\href{https://arxiv.org/abs/0907.3674}{{\ttfamily 0907.3674}}].

\bibitem{Pantaleone:1994ns}
J.T.~Pantaleone, \emph{{Neutrino flavor evolution near a supernova's core}}, \href{https://doi.org/10.1016/0370-2693(94)01369-N}{\emph{Phys. Lett. B} {\bfseries 342} (1995) 250} [\href{https://arxiv.org/abs/astro-ph/9405008}{{\ttfamily astro-ph/9405008}}].

\bibitem{Giunti:2007ry}
C.~Giunti and C.W.~Kim, \emph{{Fundamentals of Neutrino Physics and Astrophysics}} (2007).

\bibitem{randomization}
A.J.~{Majda} and P.R.~{Kramer}, \emph{{Simplified models for turbulent diffusion: Theory, numerical modelling, and physical phenomena}}, \href{https://doi.org/10.1016/S0370-1573(98)00083-0}{\emph{Physics Reports} {\bfseries 314} (1999) 237}.

\bibitem{Ohlsson:2000mj}
T.~Ohlsson, \emph{{Equivalence between neutrino oscillations and neutrino decoherence}}, \href{https://doi.org/10.1016/S0370-2693(01)00178-2}{\emph{Phys. Lett. B} {\bfseries 502} (2001) 159} [\href{https://arxiv.org/abs/hep-ph/0012272}{{\ttfamily hep-ph/0012272}}].

\bibitem{Keil:2002in}
M.T.~Keil, G.G.~Raffelt and H.-T.~Janka, \emph{{Monte Carlo study of supernova neutrino spectra formation}}, \href{https://doi.org/10.1086/375130}{\emph{Astrophys. J.} {\bfseries 590} (2003) 971} [\href{https://arxiv.org/abs/astro-ph/0208035}{{\ttfamily astro-ph/0208035}}].

\bibitem{DUNE:2020lwj}
{\scshape DUNE} collaboration, \emph{{Deep Underground Neutrino Experiment (DUNE), Far Detector Technical Design Report, Volume I Introduction to DUNE}}, \href{https://doi.org/10.1088/1748-0221/15/08/T08008}{\emph{JINST} {\bfseries 15} (2020) T08008} [\href{https://arxiv.org/abs/2002.02967}{{\ttfamily 2002.02967}}].

\bibitem{Jana:2022tsa}
S.~Jana, Y.P.~Porto-Silva and M.~Sen, \emph{{Exploiting a future galactic supernova to probe neutrino magnetic moments}}, \href{https://doi.org/10.1088/1475-7516/2022/09/079}{\emph{JCAP} {\bfseries 09} (2022) 079} [\href{https://arxiv.org/abs/2203.01950}{{\ttfamily 2203.01950}}].

\bibitem{Hyper-Kamiokande:2022smq}
{\scshape Hyper-Kamiokande} collaboration, \emph{{Hyper-Kamiokande Experiment: A Snowmass White Paper}},  in \emph{{Snowmass 2021}}, 3, 2022 [\href{https://arxiv.org/abs/2203.02029}{{\ttfamily 2203.02029}}].

\bibitem{Strumia:2003zx}
A.~Strumia and F.~Vissani, \emph{{Precise quasielastic neutrino/nucleon cross-section}}, \href{https://doi.org/10.1016/S0370-2693(03)00616-6}{\emph{Phys. Lett. B} {\bfseries 564} (2003) 42} [\href{https://arxiv.org/abs/astro-ph/0302055}{{\ttfamily astro-ph/0302055}}].

\bibitem{Mukhopadhyay:2020ubs}
M.~Mukhopadhyay, C.~Lunardini, F.X.~Timmes and K.~Zuber, \emph{{Presupernova neutrinos: directional sensitivity and prospects for progenitor identification}}, \href{https://doi.org/10.3847/1538-4357/ab99a6}{\emph{Astrophys. J.} {\bfseries 899} (2020) 153} [\href{https://arxiv.org/abs/2004.02045}{{\ttfamily 2004.02045}}].

\bibitem{Healy:2023ovi}
S.~Healy, S.~Horiuchi, M.~Colomer~Molla, D.~Milisavljevic, J.~Tseng, F.~Bergin et~al., \emph{{Red Supergiant Candidates for Multimessenger Monitoring of the Next Galactic Supernova}},  \href{https://arxiv.org/abs/2307.08785}{{\ttfamily 2307.08785}}.

\bibitem{Theia:2019non}
{\scshape Theia} collaboration, \emph{{THEIA: an advanced optical neutrino detector}}, \href{https://doi.org/10.1140/epjc/s10052-020-7977-8}{\emph{Eur. Phys. J. C} {\bfseries 80} (2020) 416} [\href{https://arxiv.org/abs/1911.03501}{{\ttfamily 1911.03501}}].

\end{thebibliography}\endgroup
\bibliographystyle{jhep}
\end{document}